\documentclass[aps,prl,showpacs,superscriptaddress,twocolumn,longbibliography]{revtex4-1}

\usepackage{amsfonts}
\usepackage{subfigure}
\usepackage{amsmath}
\usepackage{amssymb}
\usepackage{amsbsy} 
\usepackage{epsfig}
\usepackage{graphicx}
\usepackage{epstopdf}
\usepackage{color}

\def\Im{\rm{Im}}

\def\be{\begin{equation}} \def\ee{\end{equation}}
\def\bea{\begin{eqnarray}} \def\eea{\end{eqnarray}}
\def\nn{\nonumber}

\def\rw{\rightarrow}

\def\ra{\rangle}
\def\la{\langle}

\begin{document}

\title{Non-Hermitian Edge Burst}

\author{Wen-Tan Xue}
 \affiliation{ Institute for
Advanced Study, Tsinghua University, Beijing,  100084, China }
 \author{Yu-Min Hu}
 \affiliation{ Institute for
Advanced Study, Tsinghua University, Beijing,  100084, China }
\author{Fei Song}
 \altaffiliation{ songf18@mails.tsinghua.edu.cn }
\affiliation{ Institute for
Advanced Study, Tsinghua University, Beijing,  100084, China }
\author{Zhong Wang} \altaffiliation{ wangzhongemail@tsinghua.edu.cn }
\affiliation{ Institute for
Advanced Study, Tsinghua University, Beijing,  100084, China }

\begin{abstract}
We unveil an unexpected non-Hermitian phenomenon, dubbed edge burst, in non-Hermitian quantum dynamics. Specifically, in a class of non-Hermitian quantum walk in periodic lattices with open boundary condition, an exceptionally large portion of loss occurs at the system boundary. The physical origin of this edge burst is found to be an interplay between two unique non-Hermitian phenomena: non-Hermitian skin effect and imaginary gap closing. Furthermore, we establish a universal bulk-edge scaling relation underlying the non-Hermitian edge burst. Our predictions are experimentally accessible in various non-Hermitian systems including quantum-optical and cold-atom platforms.
\end{abstract}

\maketitle

Standard quantum mechanics postulates Hermiticity of Hamiltonian, yet non-Hermitian Hamiltonians are useful in many branches of physics. For example, open systems with gain and loss naturally exhibit non-Hermitian physics \cite{Ashida2021}. Recently, there have been growing interests in non-Hermitian topological physics. In particular, the bulk-boundary correspondence principle is drastically reshaped by the non-Hermitian skin effect (NHSE), namely the boundary localization of bulk-band eigenstates \cite{yao2018edge,yao2018chern,kunst2018biorthogonal,
lee2018anatomy,alvarez2017,Helbig2019NHSE,Xiao2019NHSE, Ghatak2019NHSE,Budich2020,Bergholtz2021RMP}. It indicates that the boundary plays an even more profound role in non-Hermitian systems compared to their Hermitian counterparts.

\begin{figure}
\includegraphics[width=8.5cm, height=3.2cm]{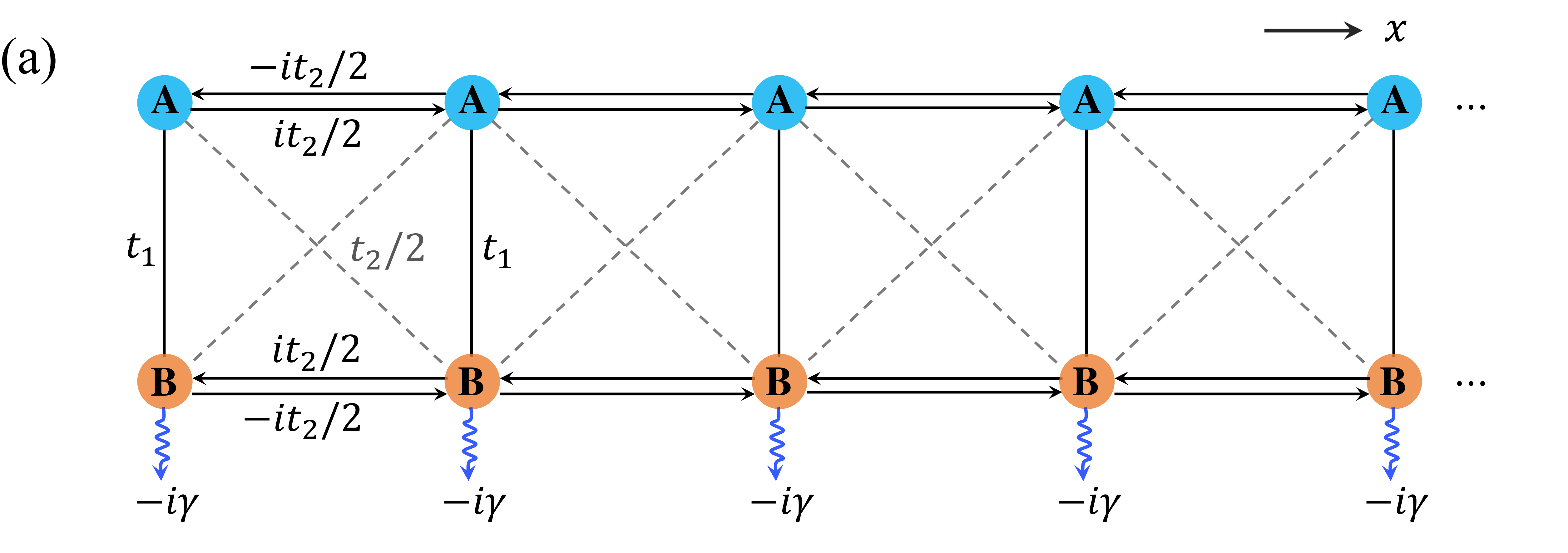}
\includegraphics[width=4.2cm, height=4cm]{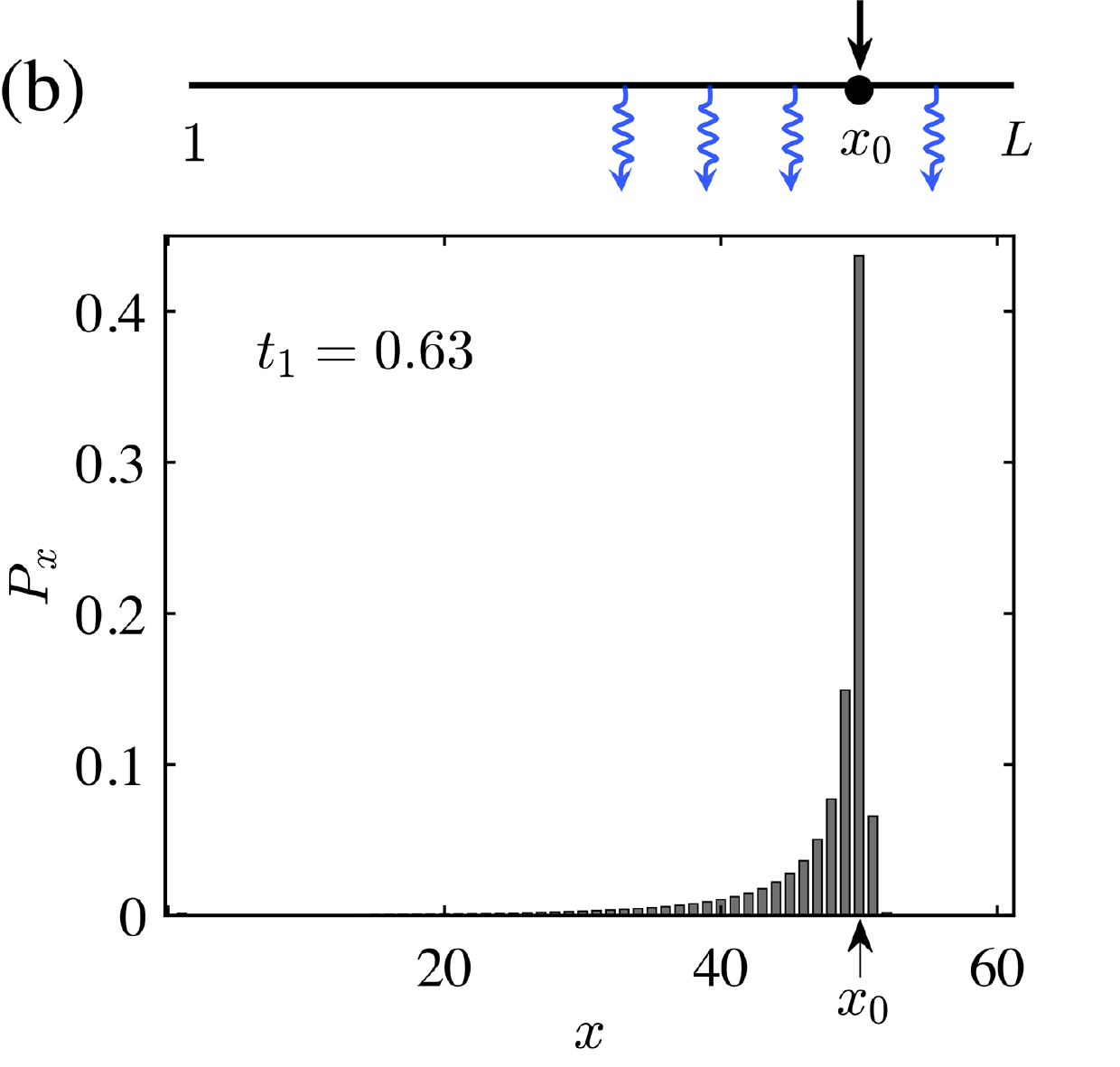}
\includegraphics[width=4.2cm, height=4cm]{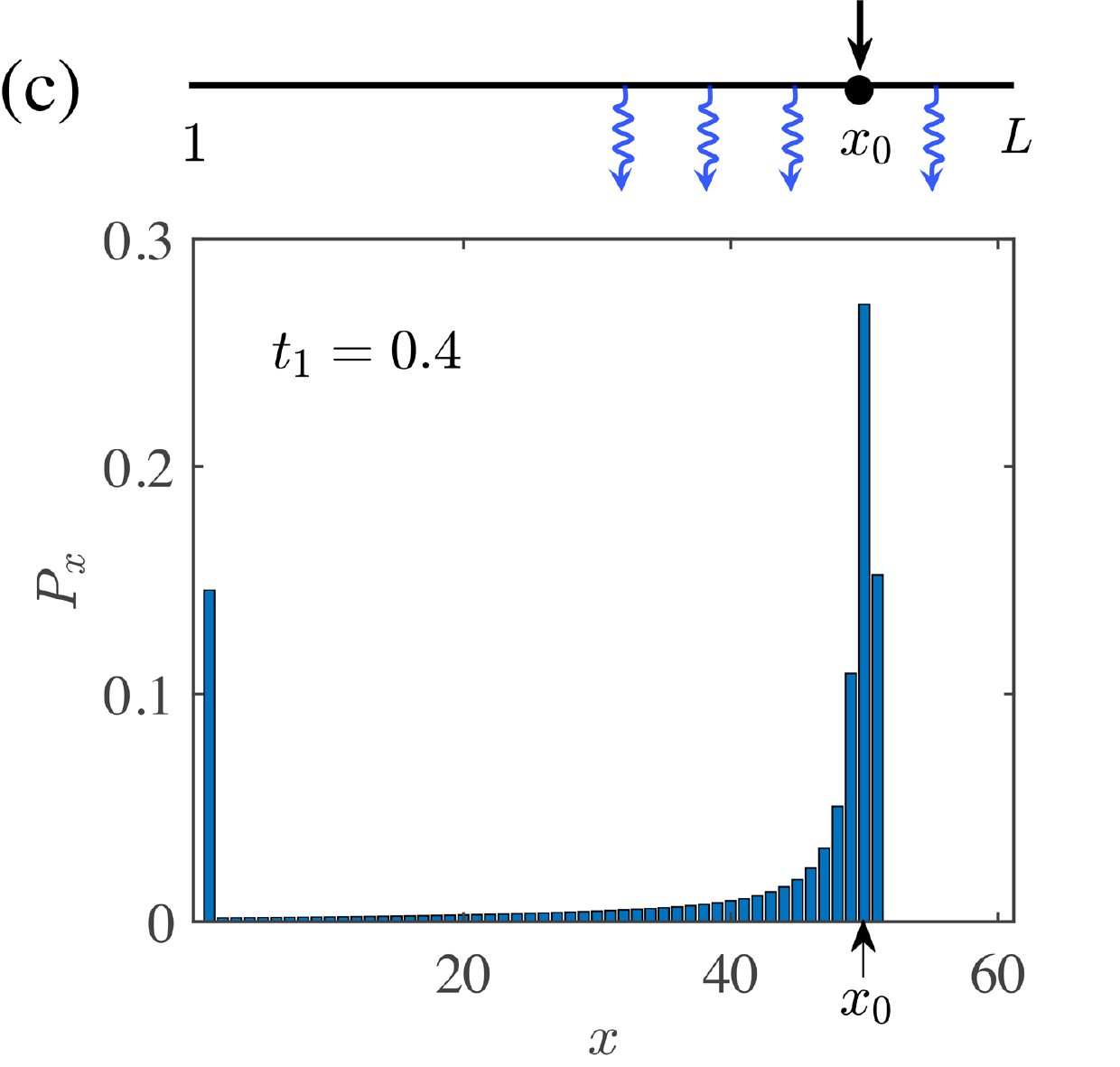}
\includegraphics[width=4.2cm, height=4cm]{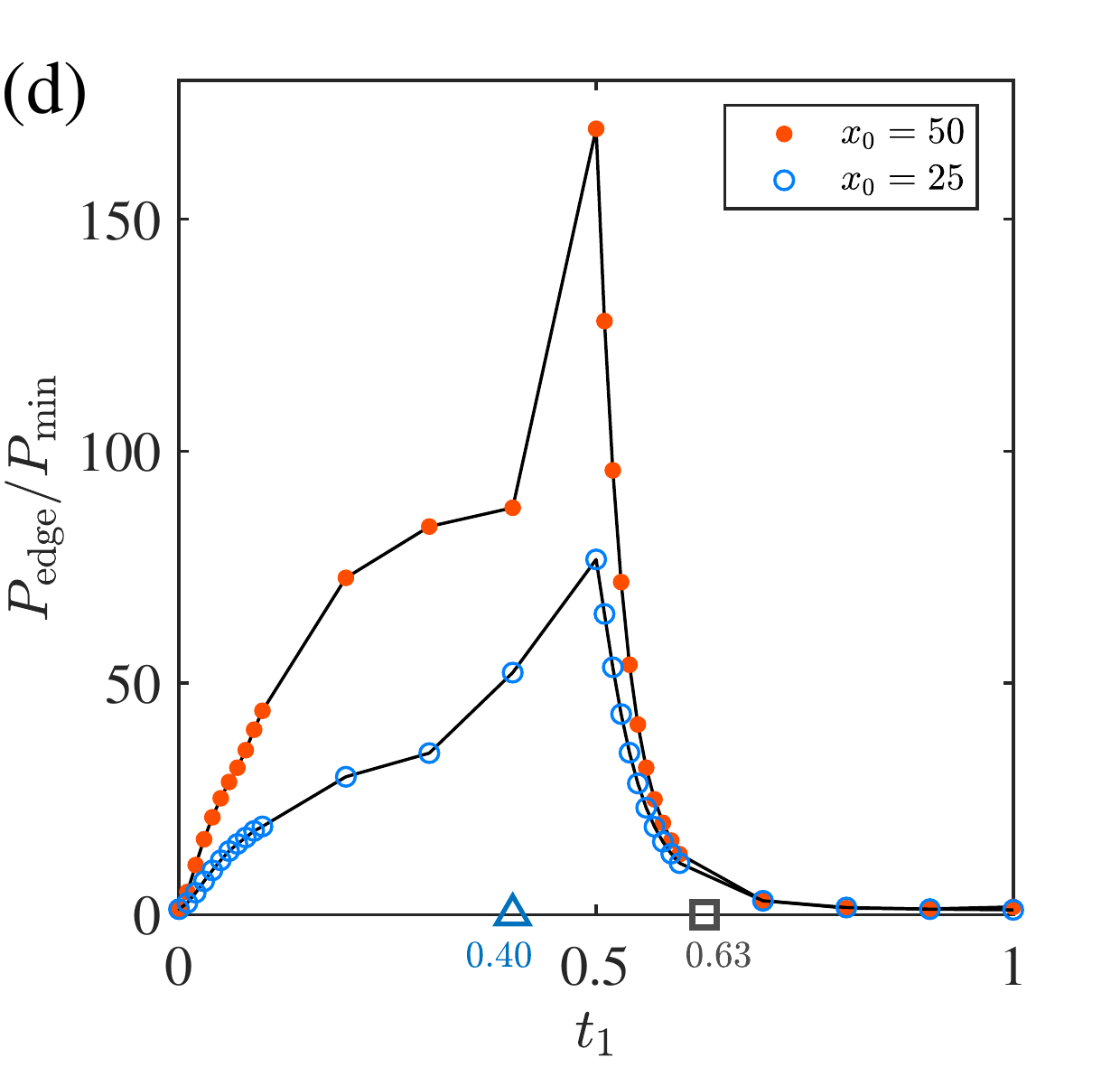}
\includegraphics[width=4.2cm, height=4cm]{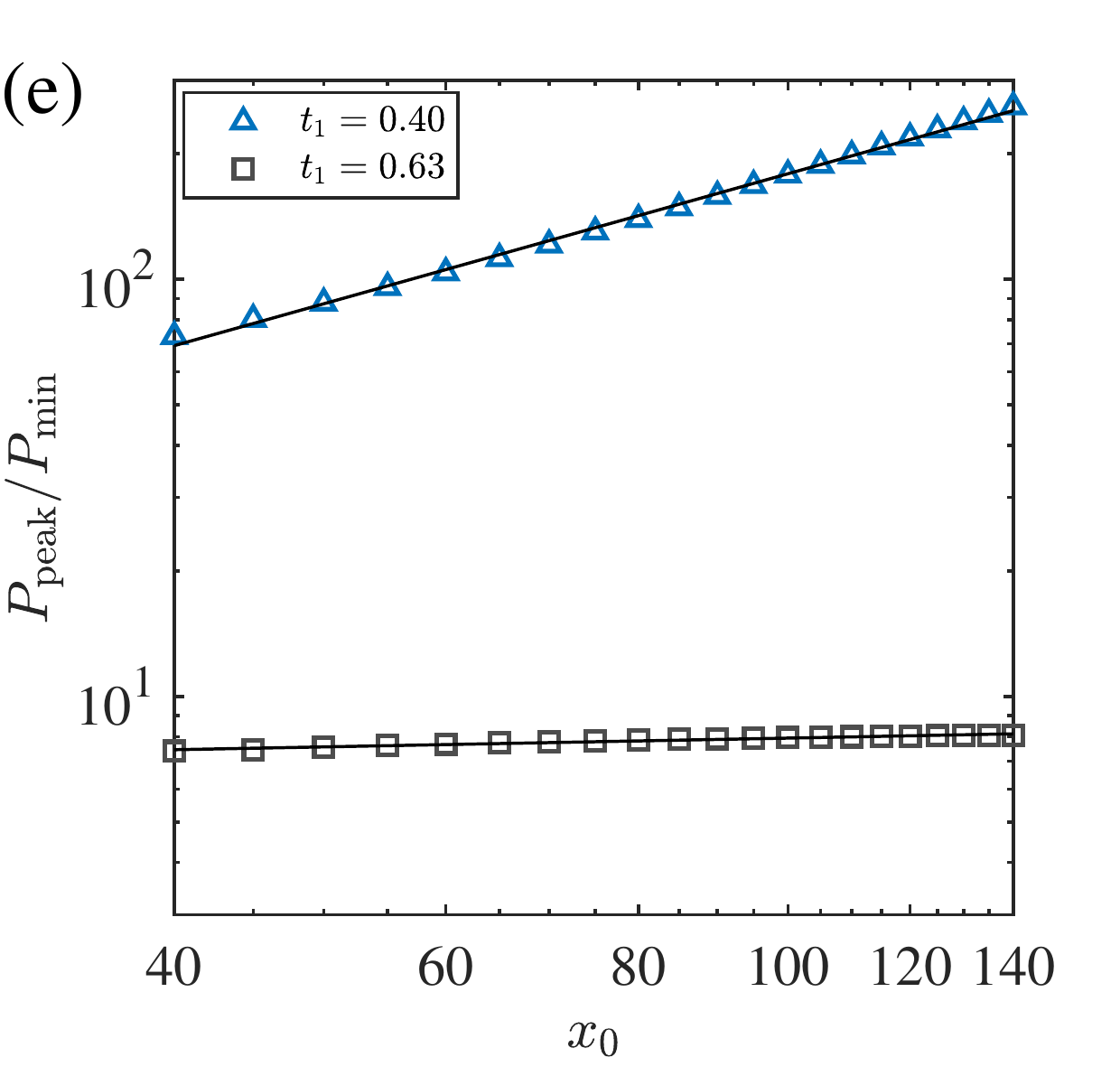}
\caption{(a) The model. Each unit cell, labeled by spatial coordinate $x$, contains two sites $A$ and $B$.  (b,c) The spatially resolved loss probability $P_x$ for a walker initiated at $x_0=50$. $t_1=0.63$ for (b) and $t_1=0.4$ for (c).  The chain length $L=60$.  (d) The relative height $P_{\rm{edge}}/P_{\rm{min}}$ with varying $t_1$, for $x_0=50$ and $25$. Here, $P_{\rm{edge}}\equiv P_{x=1}$ and $P_{\rm{min}}\equiv \min\{P_1,P_2,\cdots,P_{x_0}\}$.
(e) Relative height with $x_0$ varying from $40$ to $140$, for $t_1=0.63$ (black square) and $t_1=0.40$ (blue triangle) (marked in (d)). $L=150$.  Throughout (b-e), $t_2=0.5,\gamma=0.8$ are fixed.}
\label{fig1}
\end{figure}

In this paper, we unveil a boundary-induced dynamical phenomenon, dubbed ``edge burst'', in a class of non-Hermitian systems. For concreteness, we consider quantum-mechanical time evolution of particles (called ``quantum walkers'') in a lossy lattice. Intuitively, a walker starting from a certain site far from the edges is expected to escape predominantly from nearby sites. However, a prominent peak in the loss probability is found at the edge. More unexpectedly, the relative height of this peak grows with the distance from the initial site to the edge. Furthermore, we find that this edge burst exhibits a unique scaling behavior, originating from a universal bulk-edge scaling relation. This provides an underlying theory that not only tells the precise conditions for edge burst, but also has implications beyond.

We note that the appearance of an edge peak has been reported in a very recent work, though it was incorrectly attributed to topological edge states \cite{wang2021quantum}. Our work demonstrates that the edge burst stems entirely from the non-Bloch bulk bands, highlighting it as a robust phenomenon insensitive to edge perturbations.

\emph{Non-Hermitian edge burst.--}For concreteness, we consider a one-dimensional lossy lattice shown in Fig. \ref{fig1}(a). During the quantum walk, the walker can escape from $B$ sites. The Schr$\ddot{\text{o}}$dinger equation $i\frac{d}{dt}|\psi(t)\ra=H|\psi(t)\ra$ reads
\begin{align}
i\frac{d\psi_x^A}{dt}=&t_1\psi_x^B+i\frac{t_2}{2}(\psi_{x-1}^A-\psi_{x+1}^A)+\frac{t_2}{2}(\psi_{x-1}^B+\psi_{x+1}^B),\nn\\
i\frac{d\psi_x^B}{dt}=&t_1\psi_x^A-i\frac{t_2}{2}(\psi_{x-1}^B-\psi_{x+1}^B)+\frac{t_2}{2}(\psi_{x-1}^A+\psi_{x+1}^A)\nn\\&\qquad\qquad\qquad\qquad\qquad\qquad\quad    -i\gamma\psi_x^B, \label{realspace}
\end{align} with loss rate $\gamma>0$. The corresponding Bloch Hamiltonian is
\bea
H(k)=(t_1+t_2\cos k)\sigma_x+(t_2 \sin k+i\frac{\gamma}{2})\sigma_z-i\frac{\gamma}{2} I, \label{model1}
\eea where $\sigma_{x,y,z}$ are the Pauli matrices, with $\sigma_z= 1 (-1)$ corresponding to $A (B)$ sublattice, and $I$ is the identity matrix.  This model is similar to that of Ref. \cite{lee2016anomalous}, except that it is purely lossy. Notably, it features the NHSE, which distinguishes it from earlier quantum-walk models \cite{rudner2009topological}.  Intuitively, the $-\pi/2$ fluxes in the triangles generate rotational motions, such that the $A$ and $B$ chains favor opposite directions of motion; the loss then generates a net chiral motion along the $A$ chain by suppressing the backflow on the $B$ chain.  Alternatively, the NHSE can be seen via the equivalence of the
model, under a basis change, to the non-Hermitian Su-Schrieffer-Heeger model with left-right asymmetric hopping \cite{yao2018edge}.

The wavefunction norm decreases as
$\frac{d}{dt}\la\psi(t)|\psi(t)\ra=i\la\psi(t)|(H^\dag-H)|\psi(t)\ra=-\sum_x 2\gamma |\psi_x^{B}(t)|^2$, and the probability that the walker escapes from location $x$ is \bea P_x=2\gamma \int_0^\infty dt |\psi_x^{B}(t)|^2. \label{Px} \eea Note that $\sum_x P_x=1$ is satisfied under the initial-state normalization $\la\psi(0)|\psi(0)\ra=1$. Let us consider a walker starting from $x=x_0$, with $\psi_x^A=\delta_{x,x_0}$ and $\psi_x^B=0$. It appears natural to expect that $P_x$ would decay away from $x_0$, which is confirmed by numerical simulations [Fig. \ref{fig1}(b)]. We also notice that the $P_x$ distribution is left-right asymmetric. The preference of walking left can be attributed to the NHSE, all eigenstates being localized at the left edge \cite{yao2018edge}.

The most intriguing feature is the exceptionally high peak at the left edge, namely the edge burst, which stands out from the almost invisible decaying tail [Fig. \ref{fig1}(c)].  Such a peak was numerically seen in Ref. \cite{wang2021quantum}. However, it was unclear when and why the peak occurs. It was attributed to topological edge states, which turns out to be incorrect. In fact, both (b) and (c) in Fig. \ref{fig1} are within the topologically nontrivial regimes (i.e. there are topological edge modes) \cite{yao2018edge,kunst2018biorthogonal}, yet the edge burst occurs only in (c), which looks puzzling.

To quantify the edge burst, we calculate the relative height, defined as $P_\text{edge}/P_\text{min}$, where $P_\text{edge}=P_1$, and $P_{\rm{min}}\equiv \min\{P_1, P_2,\cdots,P_{x_0}\}$ is the minimum of $P$ between the starting point and the edge. The existence and absence of edge burst manifests in $P_\text{edge}/P_\text{min}\gg 1$ and $P_\text{edge}/P_\text{min}\sim 1$, respectively. We see in Fig. 1(d) that the relative height increases with $x_0$ for $t_1\in (0, t_2]$ (approximately), and rapidly decreases to order of unity otherwise, with $t_2=0.5$ fixed. In Fig. 1(e), we plot the relative height for $t_1=0.40$ and $0.63$, which grows with $x_0$ in the former case.  The numerical fitting $P_\text{edge}/P_\text{min}\sim (x_0)^{1.03}$ is close to being linear. We note that NHSE is present for all $t_1\neq 0$, and therefore Fig. 1(d)(e) tell us that NHSE by itself does not guarantee edge burst.

\begin{figure}
\includegraphics[width=4.2cm, height=4cm]{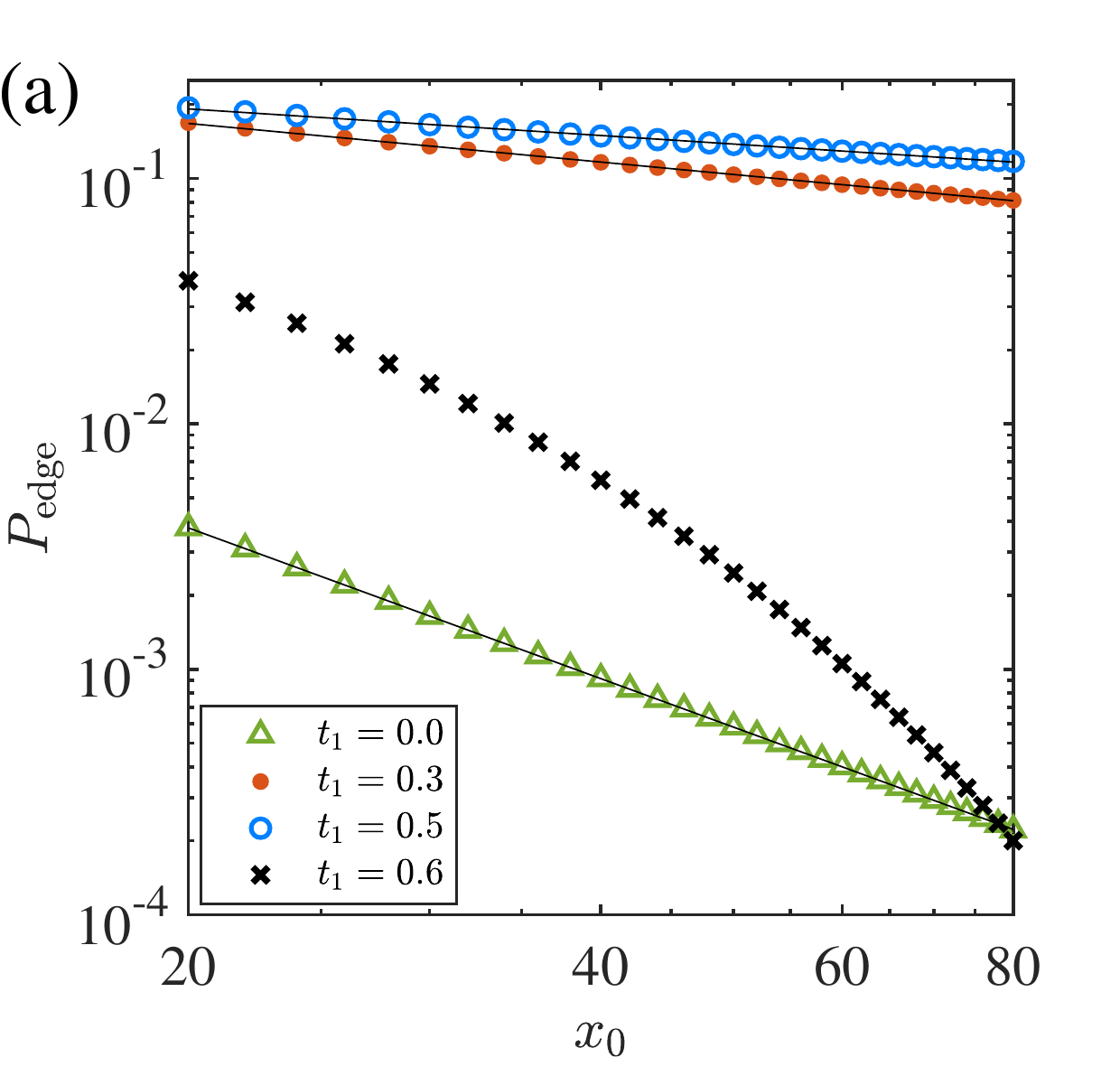}
\includegraphics[width=4.2cm, height=4cm]{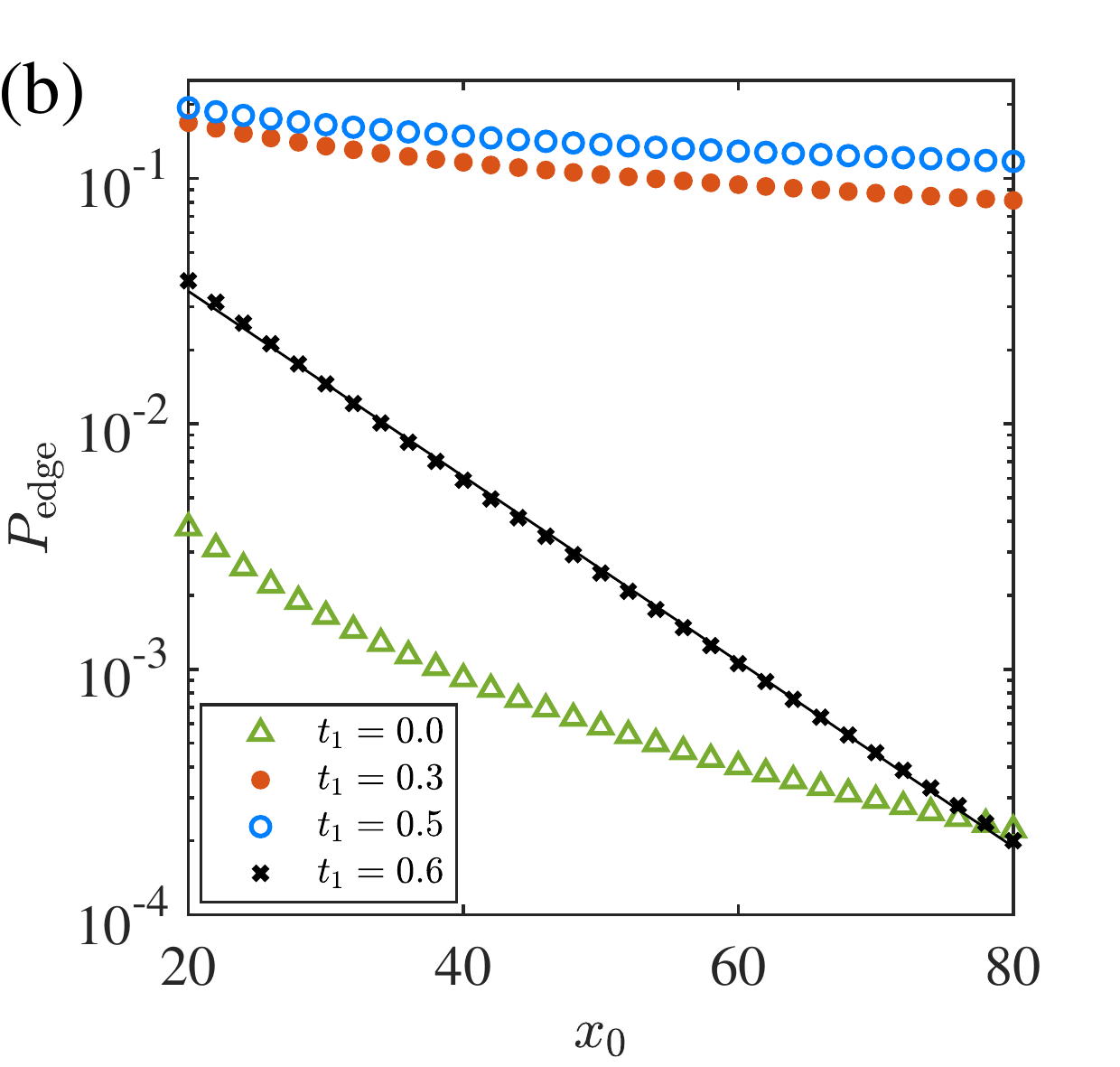}
\includegraphics[width=4.2cm, height=4cm]{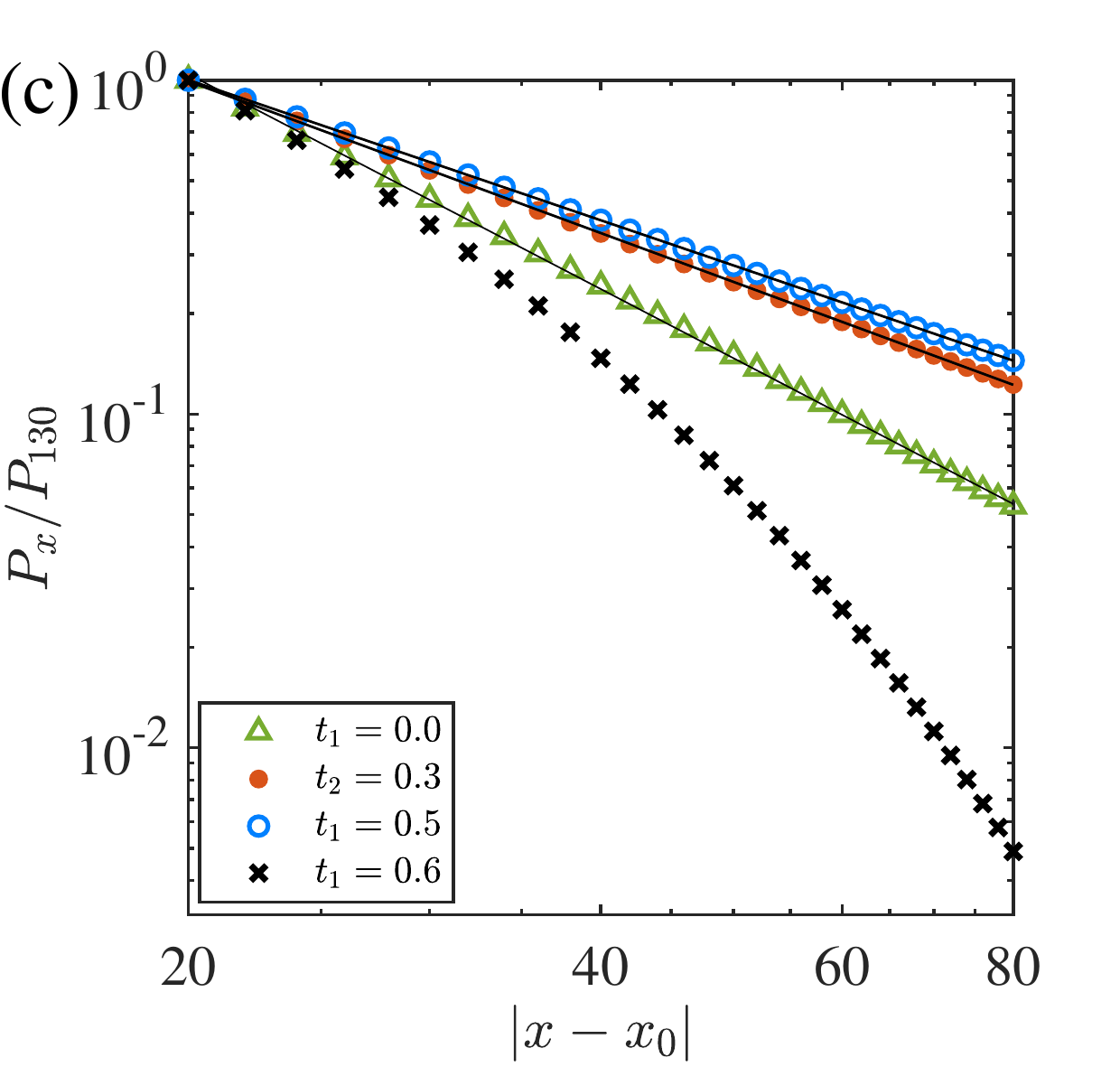}
\includegraphics[width=4.2cm, height=4cm]{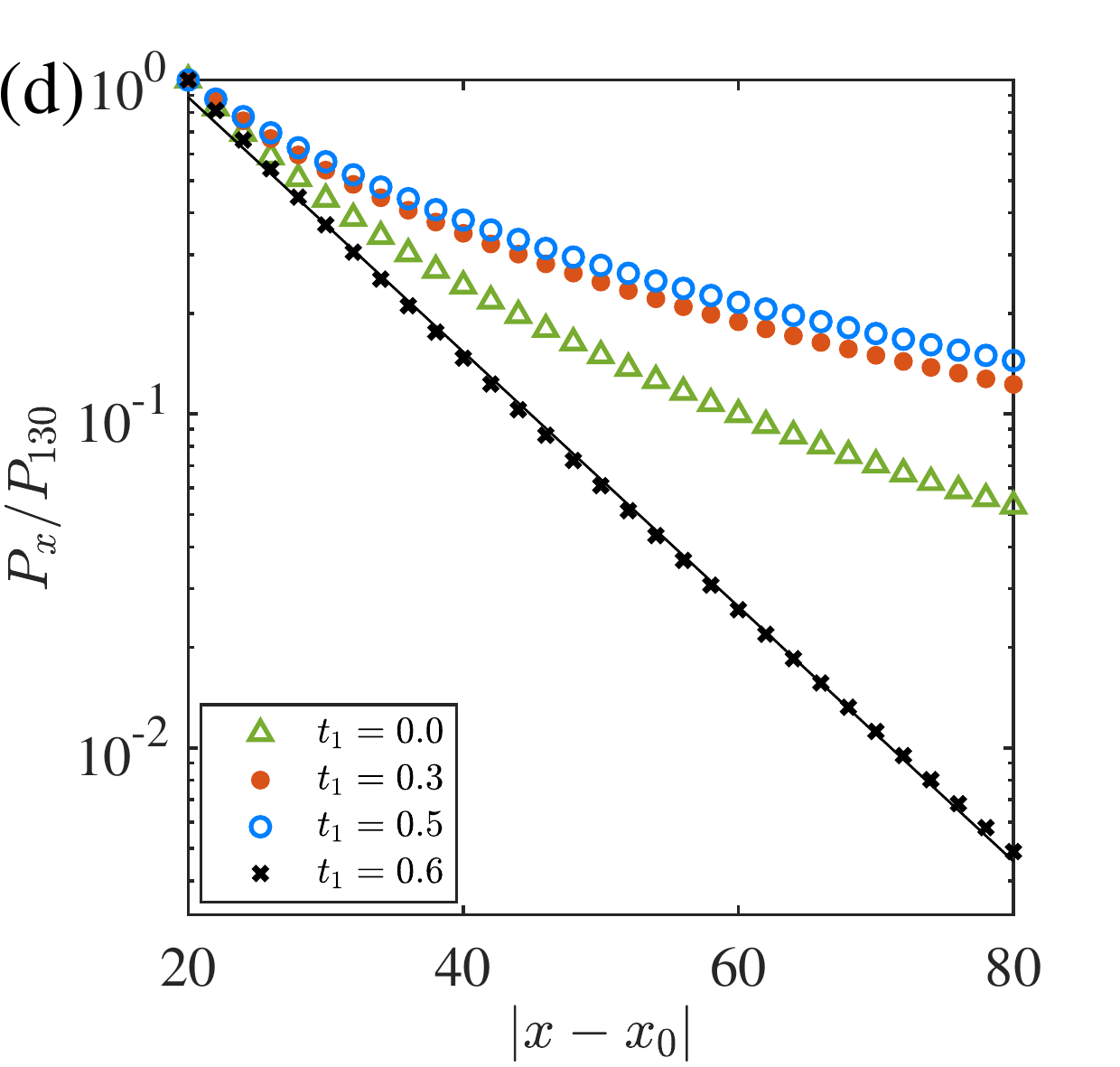}
\includegraphics[width=4.2cm, height=4cm]{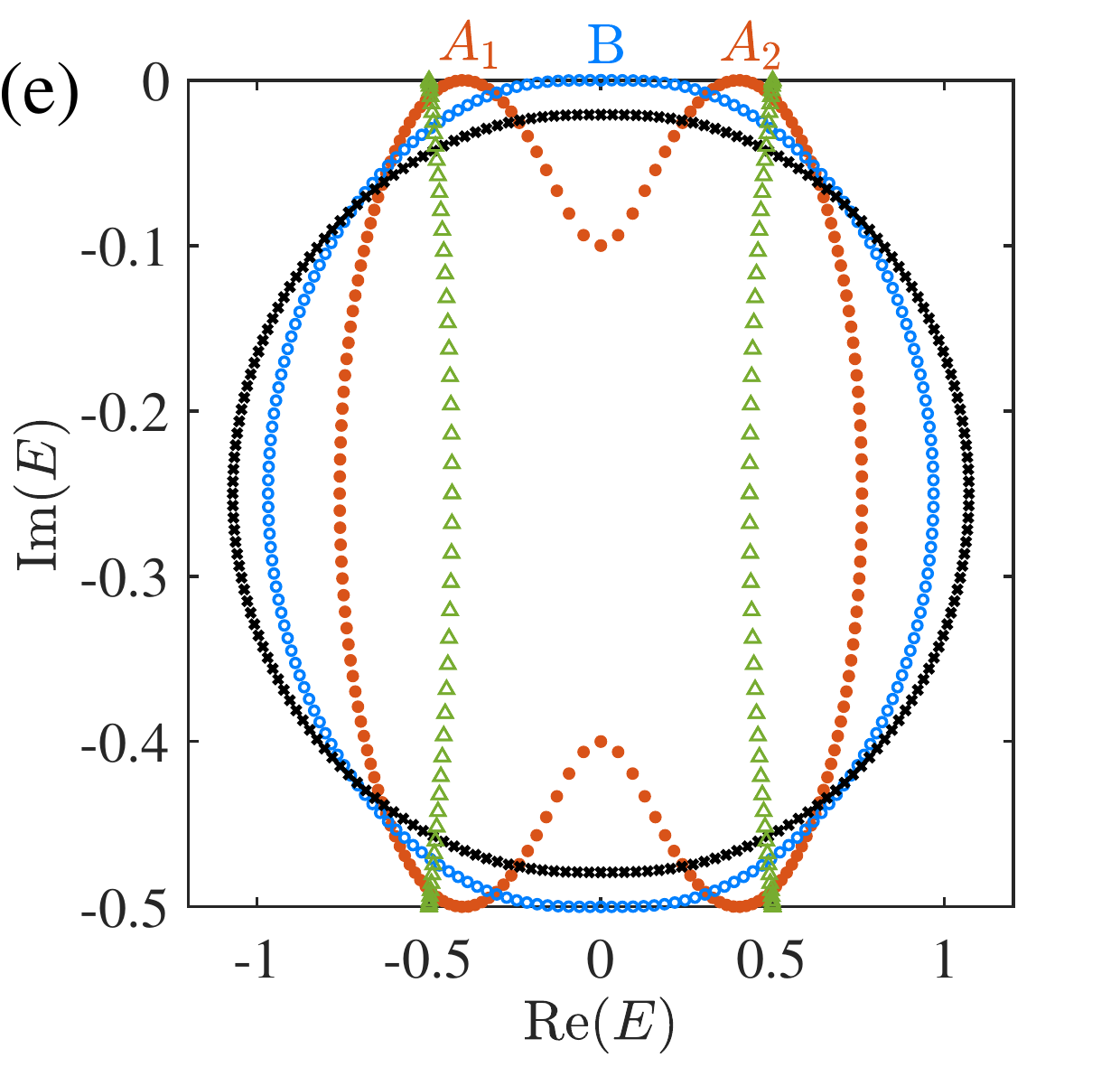}
\includegraphics[width=4.2cm, height=4cm]{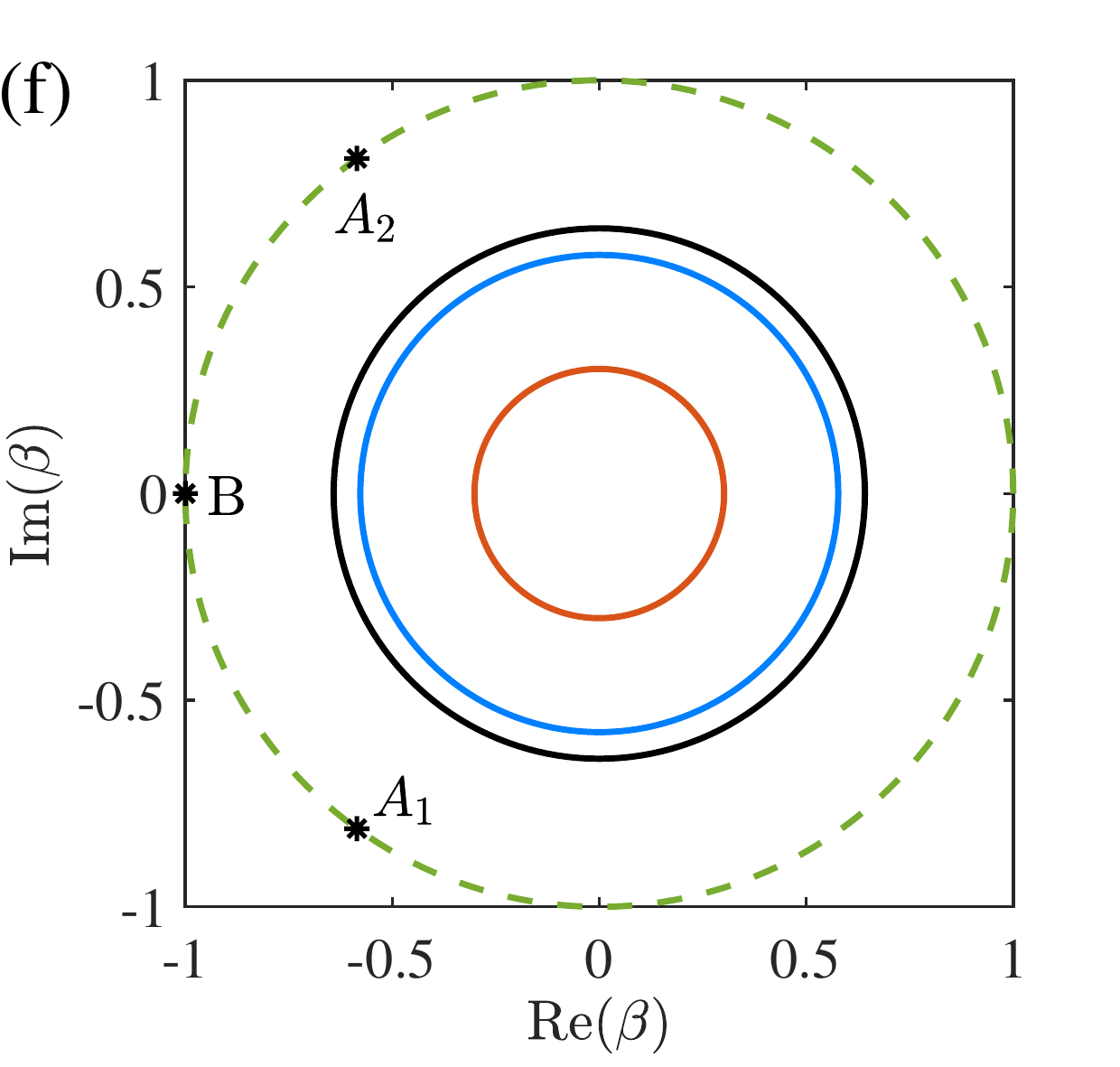}
\caption{(a)(b) The height of edge peak in double logarithmic (a) and logarithmic (b) plot. (c)(d) The bulk distribution of $P_x$ in double logarithmic (c) and logarithmic (d) plot. $L=200$ for (a-d), and $x_0=150$ for (c,d).  (e) Energy spectrums under periodic boundary condition (PBC). The green, red, and blue spectrums close the imaginary gap (dissipative gap), i.e. touch the real axis, while the black spectrum exhibits a nonzero imaginary gap. (f) Generalized Brillouin zone (GBZ). Throughout (a-f), $t_2=0.5,\gamma=0.5$, $t_1=0$ (green), $0.3$ (red), $0.5$ (blue), and $0.6$ (black).  }
\label{fig2}
\end{figure}

To unveil the origin of edge burst, we plot both $P_\text{edge}$ and bulk $P_x$ in Fig. 2. Fig. 2 (a,b) indicate that $P_\text{edge}$ follows a power law for $|t_1|\leq |t_2|$, \bea P_\text{edge}\sim |x_0|^{-\alpha_e}, \label{edge} \eea and an exponential law $P_\text{edge}\sim (\lambda_e)^{x_0}$ for $|t_1|> |t_2|$. Fig. 2 (c,d) indicate similar behaviors in the bulk, \bea P_x\sim |x-x_0|^{-\alpha_b}, \label{bulk} \eea for $|t_1|\leq  |t_2|$, and exponential law $P_x\sim (\lambda_b)^{x_0-x}$ ($\lambda_b<1$) for $|t_1|>|t_2|$. Note that Eq. (\ref{bulk}) is valid only for $x$ in the bulk, i.e. not too close to the edge; also note that $\alpha_b\neq \alpha_e$. The algebraic (i.e. power-law) behavior of bulk $P_x$ reflects the algebraic decay of wavefunction norm in the time domain, which originates from the Bloch energy spectrum touching the real axis, i.e. closing the imaginary gap [Fig. 2(e)]. In other words, algebraic decay corresponds to $\text{max}[\text{Im}E(k)]=0$, with $E$ denoting the eigen-spectrums of $H$. It can be readily checked that the imaginary gap closes for $|t_1|\leq |t_2|$ \cite{supplemental}.

In the language of open quantum system, the algebraic behavior means that the dissipative gap (or Liouvillian gap) closes \cite{Cai2013algebraic}. In fact, our non-Hermitian $H$ in Eq. (\ref{realspace}) can be reformulated in terms of the quantum master equation, $\frac{d\rho}{dt}=-i[\mathcal{H},\rho]+\sum_x(L_x\rho L_x^\dag -\frac{1}{2}\{L^\dag_x L_x,\rho\})$, where $\mathcal{H}=\sum_{i,j} c^\dag_i h_{ij} c_j$, with $h$ denoting the Hermitian part of $H$ in Fig. 1(a), namely, $h_{ij}=H_{ij}(\gamma=0)$, and the dissipator $L_x=\sqrt{2\gamma}c_x^B$. Note that $c_i$ can be either bosonic or fermionic, which does not affect the single-particle dynamics.  The effective non-Hermitian Hamiltonian $H_\text{eff}= \mathcal{H}-\sum_x \frac{1}{2} L^\dag_x L_x = \mathcal{H}-\gamma\sum_x c^{B\dag}_x  c_x^B = \sum_{ij} c^\dag_i H_{ij} c_j$. In this context, $\text{max}[\text{Im}E(k)]=0$ corresponds to closing the dissipative (imaginary) gap.

Given the imaginary gap closing, namely $|t_1|\leq |t_2|$, we always see the edge burst except at $t_1=0$. The $t_1=0$ point is special in two aspects. First, NHSE is absent at this parameter value. Second, the periodic-boundary-condition (PBC) energy spectrum encloses zero area in complex plane [green triangle in Fig. 2(e)]. These two features are concurrent. In fact, a precise correspondence has been established between the existence (absence) of NHSE and the complex energy enclosing nonzero (zero) area \cite{Zhang2020correspondence,Okuma2020}. The zero and nonzero enclosed area is also known as having trivial and nontrivial point-gap topology, respectively \cite{kawabata2019symmetry,gong2018nonhermitian,shen2017topological}.

Summarizing the above numerical findings, we infer that the edge burst stems from the interplay between two prominent non-Hermitian phenomena, NHSE and imaginary gap closing. The latter is a non-Hermitian counterpart of being gapless in Hermitian systems. This imaginary gaplessness and NHSE jointly induce the edge burst.

\emph{Bulk-edge scaling relation.--}The exponent $\alpha_e$ in Eq. (\ref{edge}) and $\alpha_b$ in Eq. (\ref{bulk}) characterize the edge and bulk dynamics, respectively. One of our central results is the scaling relation
\bea
\alpha_e=\alpha_b-1  \label{hyperscaling}
\eea
in the presence of NHSE and imaginary gap closing. For our specific model, it holds true when $|t_1|\leq |t_2|$ (such that imaginary gap closes) and $t_1\neq 0$ (such that NHSE is present). At the NHSE-free point $t_1=0$, we have $\alpha_e=\alpha_b$ instead. Remarkably, although both $\alpha_b$ and $\alpha_e$ are model/parameter dependent, the relation Eq. (\ref{hyperscaling}) remains universal.  Numerical fitting in Fig. \ref{fig2}(a)(c) yields $\alpha_b-\alpha_e=0.99, 1.03$ for $t_1=0.3, 0.5$, respectively, which is in reasonable agreement with Eq. (\ref{hyperscaling}). For $t_1=0$, the fitting yields $\alpha_b-\alpha_e=0.09$, being close to the theoretical value $0$ for the NHSE-free cases.

Before calculating $\alpha_b, \alpha_e$ and proving Eq. (\ref{hyperscaling}), we observe that this equation implies edge burst. In fact, Eq. (\ref{bulk}) implies that $P_x$ takes the minimum near (but not too close to) the edge, and $P_\text{min}\sim x_0^{-\alpha_b}$. Therefore, it follows from Eq. (\ref{hyperscaling}) that \bea  P_\text{edge}/P_\text{min} \sim  x_0^{\alpha_b-\alpha_e} \sim x_0. \label{burst} \eea
Thus, as the starting point $x_0$ moves away from the edge, the relative height of edge peak increases. This is precisely the origin of edge burst.

Now we calculate $\alpha_b, \alpha_e$ and derive Eq. (\ref{hyperscaling}) using Green's function, which has been a useful tool in non-Hermitian systems \cite{McDonald2018phase,Xue2021simple,Wanjura2019,carlstrom2020correlations,zirnstein2021,Borgnia2019}. The integrand in Eq. (\ref{Px}) can be expressed as $|\la x,B|G(t)|\psi(t=0)\ra|^2$, where $G(t)=-i\Theta(t)e^{-iHt}$, with $\Theta(t)$ standing for the Heaviside step function.  It is convenient to work in the frequency (energy) domain using $G(t)=\frac{1}{2\pi}\int_{-\infty}^{+\infty} d\omega G(\omega) e^{-i\omega t}$, in which the Green's function reads $G(\omega)=\frac{1} {\omega+i0^{+}-H}$. Now we can recast Eq. (\ref{Px}) into
\bea
P_x=\frac{\gamma}{\pi}\int_{-\infty}^{+\infty} d\omega |\la x,B|G(\omega)|x_0,A\ra|^2.
\eea
where the initial state $|\psi(t=0)\ra=|x_0,A\ra$ has been inserted. To calculate $\alpha_b$, it is more convenient to consider an \emph{infinite chain}. The relevant Green's function reads
\begin{align}
\la x,B|G(\omega)|x_0,A\ra&=\int_0^{2\pi}\frac{dk}{2\pi}  e^{ik(x-x_0)}\left(\frac{1}{\omega+i0^{+}-H(k)}\right)_{BA}\nn\\
                  &=\oint_{|\beta|=1} \frac{d\beta}{2\pi i\beta} \beta^{x-x_0} \left(\frac{1}{\omega+i0^{+}-H(\beta)}\right)_{BA},\label{toeplitz}
\end{align} where $H(\beta)$ is the analytic continuation of $H(k)$ in Eq~(\ref{model1}), $H(\beta)\equiv H(k)|_{e^{ik}\rw\beta}$. For our specific model, $(\frac{1}{\omega+i0^{+}-H(\beta)})_{BA}=(t_1+t_2\frac{\beta+\beta^{-1}}{2})/\det[\omega+i0^{+}-H(\beta)]$. This integration can be done by the residue theorem, and the asymptotic behavior at $|x-x_0|\rw\infty$ is determined by the roots of $\det[\omega+i0^{+}-H(\beta)]=0$ \cite{Xue2021simple}. As a quadratic equation, it has two roots that we order as $|\beta_{L}(\omega)|\geq |\beta_{R}(\omega)|$. Following Ref. \cite{Xue2021simple}, we have $\la x,B|G(\omega)|x_0, A\ra\sim f_L\beta_L^{x-x_0}$ for $x<x_0$,  and $\la x, B|G(\omega)|x_0,A\ra\sim f_R\beta_R^{x-x_0}$ for $x>x_0$ ($|\beta_L(\omega)|\geq 1\geq |\beta_R(\omega)|$ is satisfied for real-valued $\omega$), where $f_{L/R}$ are $x$-independent and their precise values do not concern us \cite{supplemental}. Accordingly, $P_x^{\infty}$, in which the superscript $\infty$ stands for the infinite chain, is given by
\bea
P_x^{\infty}=\frac{\gamma}{\pi}\int_{-\infty}^{+\infty}d\omega|f_{L/R}(\omega)|^2|\beta_{L/R}(\omega)|^{2(x-x_0)}, \label{integral}
\eea
where the subscript $L$ and $R$ corresponds to $x<x_0$ and $x>x_0$, respectively. For $|x-x_0|$ large, the integral of Eq. (\ref{integral}) is dominated by $|\beta|$ closest to $1$.
In fact, the existence (absence) of a real $\omega$ satisfying $|\beta(\omega)|=1$ determines the algebraic (exponential) behavior of $P_x$. To satisfy $|\beta(\omega)|=1$ for real-valued $\omega$ is to close the imaginary gap of Bloch Hamiltonian, because the gap-closing point $\omega_0$ satisfies $\det[\omega_0-H(\beta)]=0$ with $|\beta|=1$.  These $\omega_0$ values are marked as $A_1$, $A_2$ and $B$ in Fig.~\ref{fig2}(e), and the corresponding $\beta$ values in Fig.~\ref{fig2}(f).

As we focus on $x<x_0$, the relevant root is $\beta_L(\omega)$. Let us write $\omega=\omega_0+\delta\omega$, and then expand $\beta_L(\omega), f_L(\omega)$ to the lowest order of $\delta\omega$, so that $|\beta_L(\omega)|\approx 1+K\delta\omega^n\approx \exp(K\delta\omega^n)$, and $|f(\omega)|^2\sim \delta\omega^m$. Now $P_x^{\infty} \sim \int d(\delta\omega)  \delta\omega^m \exp(-2K\delta\omega^n|x-x_0|) \sim |x-x_0|^{-(m+1)/n}$, and therefore $\alpha_b=(m+1)/n$. In contrast, when the imaginary gap opens, we have $|\beta_L(\omega)|>1$ and exponential decay $P_x^{\infty}\sim [\text{min}(\beta_L(\omega))]^{-2|x-x_0|}$. For our model, the imaginary gap closing regime is $|t_1|\leq |t_2|$, in which the bulk $P_x$ indeed exhibits algebraic behavior. Furthermore, taking $t_2>0$, we have $(n,m)=(1/2,0)$, $(2,2)$, and $(4,4)$ for $t_1=0$, $t_1\in (0,t_2)$, and $t_1=t_2$, respectively. This leads to the bulk exponent \cite{supplemental} \bea \alpha_b=\frac{m+1}{n}=2, \, \frac{3}{2}, \, \frac{5}{4}, \eea for these three cases, which is in reasonable agreement with the numerical values $\alpha_b=2.13, 1.51, 1.39$ obtained from Fig. \ref{fig2}(c).

\begin{figure}
\includegraphics[width=8cm, height=4cm]{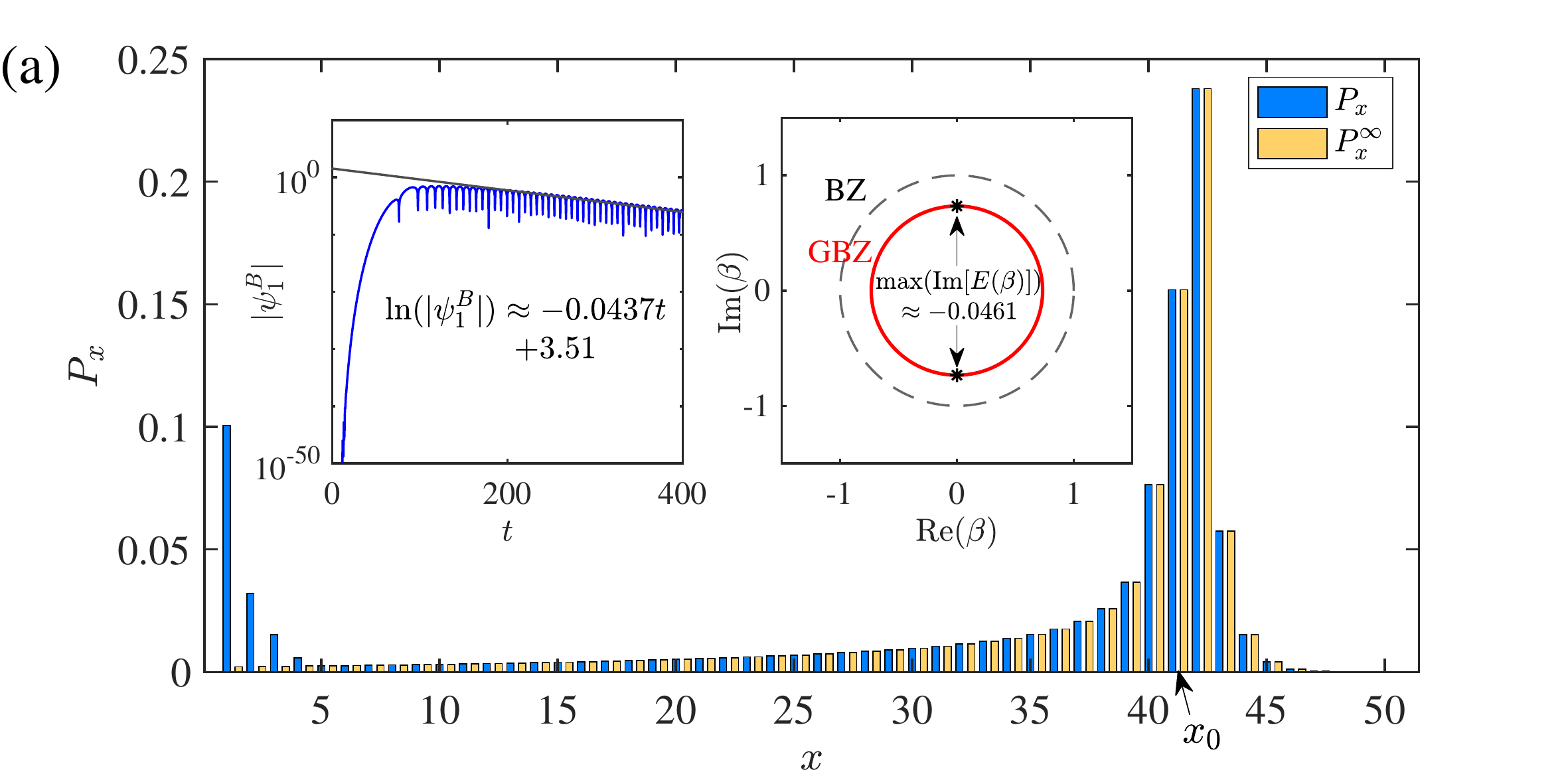}
\includegraphics[width=8cm, height=4cm]{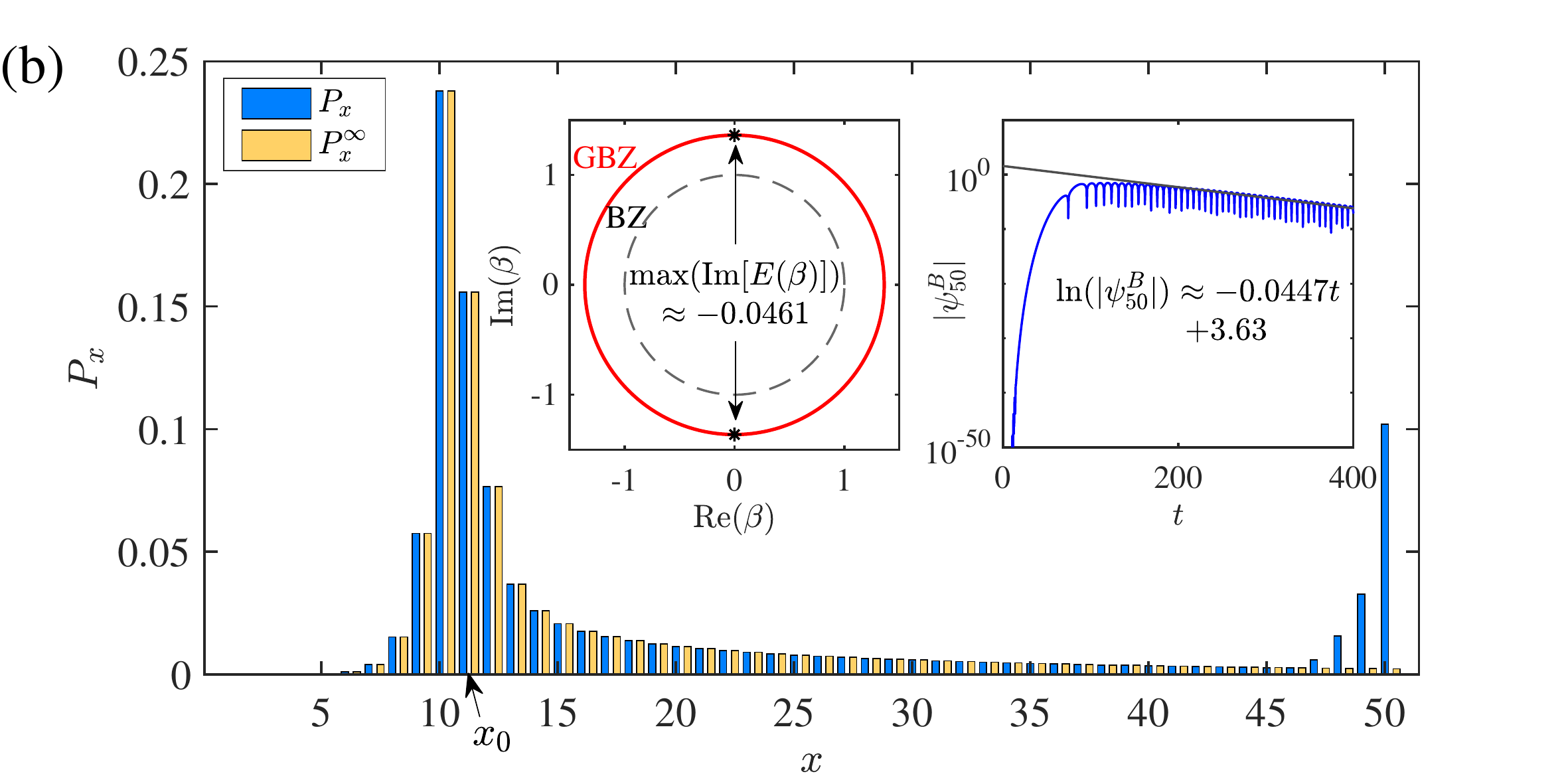}
\caption{ (a) $P_x$ for OBC chain with $L=50$ ($x=1,2\cdots 50$) (blue) and the infinite chain (yellow). The latter is represented by a $L=150$ chain ($x=-99,-98\cdots,50$), which is effectively infinite since the walker remains far from the edge throughout the time evolution. Only the $[1,50]$ interval is shown. The left inset shows the long-time evolution of the wavefunction at the edge for $L=50$. The right inset shows the GBZ. $t_2=0.5$, $\gamma=2$, $t_1=0.3$, and $x_0=41$.    (b) Similar to (a) except that $t_1=-0.3$. The infinite chain is represented by a $L=150$ chain with $x=1,2\cdots 150$, and $x_0=11$.  }
\label{fig3}
\end{figure}

Now let us consider a chain with open boundary condition (OBC) at $x=1$ and $L$ [Fig. \ref{fig1}(b)]. The NHSE of our model localizes all eigenstates exponentially to the edge. This effect can be precisely characterized by the generalized Brillouin zone (GBZ), which is the trajectory of $\beta$ associated with OBC eigenstates \cite{yao2018edge,Yokomizo2019,Longhi2019Probing,Yang2020Auxiliary, Deng2019,Kawabata2020nonBloch}. In our model, the GBZ is a circle with radius $|\beta|=\sqrt{|(t_1-\gamma/2)/(t_1+\gamma/2)|}<1$ for $t_1>0$, indicating NHSE with skin modes localized at the left edge \cite{yao2018edge}. The NHSE induces leftward walking, and the walker becomes trapped at the left edge once it arrives there.
We compare the $P_x$ of the (effectively) infinite chain and finite chain [Fig. \ref{fig3}(a)], which indicates that $P_x^\infty$ is almost the same as OBC $P_x$ for $x$ not too close to the edge. In view of the probability sum $\sum_x P_x=1$ in both cases, we conclude that the missing part, namely the edge accumulation in the OBC case and $\sum_{x=-\infty}^{0}P_x^\infty$ in the infinite-chain case, must be equal. This observation leads to the estimation
\bea P_\text{edge} &\sim&  \sum_{x=-\infty}^{0}P_x^\infty \sim \int_{-\infty}^0 |x-x_0|^{-\alpha_b}dx \nn \\ &\sim& \int_{x_0}^\infty  x^{-\alpha_b} dx\sim (x_0)^{-\alpha_b+1}. \label{key} \eea Therefore, we see that $\alpha_e$ in Eq. (\ref{edge}) equals $\alpha_b-1$. As explained by Eq. (\ref{burst}), this ``$-1$'' in exponent means a dramatic enhancement of $P_\text{edge}$ compared to the decay tail of $P_x$, generating the edge burst. In contrast, when the imaginary gap is nonzero, we have $P_\text{edge} \sim \int_{-\infty}^{0} (\lambda_b)^{x_0-x}dx \sim \int_{x_0}^{\infty} (\lambda_b)^x dx\sim (\lambda_b)^{x_0}$, which is of the same order as the decay tail (taking $x=0$ in $P_x\sim\lambda_b^{x_0-x}$), and therefore no edge burst exists. Moreover, it implies $\lambda_e=\lambda_b$.  Numerical fitting in Fig. \ref{fig2}(b)(d) yields $\lambda_b\approx 0.916$ and $\lambda_e\approx 0.917$, being close to each other.

Our calculations above demonstrate the respective role of imaginary gap closing and NHSE in creating the edge burst. The former causes the algebraic decay of $P_x$ in the bulk, while the latter drives chiral motion and contributes the crucial ``$-1$'' to the right-hand side of Eq. (\ref{hyperscaling}).

Since $\alpha_b$ is a bulk-band quantity, Eq. (\ref{key}) and Eq. (\ref{hyperscaling}) unambiguously tells that the edge bust is a bulk-band phenomenon independent of edge details. The bulk-band nature can also be seen in the long-time behavior of wavefunction. In fact, we can write $H=\sum_n E_n |n_R\ra\la n_L|$ in terms of the right and left eigenstates $|n_R\ra$ and $|n_L\ra$, then $\la x,B|\psi(t)\ra=\sum_n e^{-iE_nt}\la x, B|n_R\ra\la n_L|x_0,A\ra$. It follows that $\text{max}\{{\Im}(E_n)\}$ dominates the long-time behavior, and $|\la x,B|\psi(t)\ra|\sim e^{\max\{{\Im}(E_n)\}t}$ for $t\rightarrow \infty$. Under OBC, the bulk band consists of skin modes localized at the edge, and $E_n$ should be calculated from GBZ \cite{yao2018edge,Yokomizo2019}. According to Longhi \cite{Longhi2019Probing}, $\text{max}\{{\Im}(E)\}$ of bulk band occurs at a saddle point $\beta_s$ on GBZ, satisfying $(\partial E/\partial \beta)_{\beta=\beta_s}=0$. We numerically calculate the time dependence of edge-site wavefunction, which indeed follows an exponential law with exponent close to $\text{max}\{{\Im}(E)\}t$  [insets of Fig. \ref{fig3}(a)], confirming the bulk-band nature of edge burst. To further back up our results, we change the sign of $t_1$ so that the skin modes and edge burst are seen at the right edge; the results again support our picture [Fig. \ref{fig3}(b)]. Results from other models, including those with bipolar NHSE  \cite{Song2019real}, also confirmed our theory \cite{supplemental}.

\emph{Conclusions.--}We unveil a boundary-induced non-Hermitian dynamical phenomenon, dubbed the edge burst, which is an unexpected interplay between imaginary (dissipative) gap and NHSE. Its origin is identified as a universal bulk-edge scaling relation [Eq. (\ref{hyperscaling})]. Our theory can be readily confirmed in various non-Hermitian platforms including, for example, the photon quantum walk in which NHSE has been realized and the dissipative gap can be conveniently tuned \cite{Xiao2019NHSE,Xiao2021nonBloch}. Dissipative cold atom systems with NHSE is also a promising platform \cite{Gou2020,Lapp2019}.

{\it Acknowledgements.--} This work is supported by NSFC under Grant No. 12125405.

\bibliography{dirac}

\end{document}


\title{Supplemental Material for ``Non-Hermitian Edge Burst"}

\author{Wen-Tan Xue}
 \affiliation{ Institute for
Advanced Study, Tsinghua University, Beijing,  100084, China }
 \author{Yu-Min Hu}
 \affiliation{ Institute for
Advanced Study, Tsinghua University, Beijing,  100084, China }
\author{Fei Song}
 \altaffiliation{ songf18@mails.tsinghua.edu.cn }
\affiliation{ Institute for
Advanced Study, Tsinghua University, Beijing,  100084, China }
\author{Zhong Wang} \altaffiliation{ wangzhongemail@tsinghua.edu.cn }
\affiliation{ Institute for
Advanced Study, Tsinghua University, Beijing,  100084, China }

\maketitle
\onecolumngrid

%
%
%
%
%
%
%
%
%

\section{ Imaginary gap closing conditions }

We have stated in the main article that the imaginary gap closes for $|t_1|\leq |t_2|$. This can be checked by straightforward calculations. A less computational approach is as follows.

The model considered in our main article is reproduced as follows:
\bea
\text{Model I}:\ \ H(k)=(t_1+t_2\cos k)\sigma_x+(t_2 \sin k+i\frac{\gamma}{2})\sigma_z-i\frac{\gamma}{2} I.
\eea
For later use, we also consider a different model
\bea
\text{Model II}:\ H(k)=(t_1+t_2\cos k) \sigma_x+[t_3\cos(k-\alpha)+i\frac{\gamma}{2}]\sigma_z-i\frac{\gamma}{2}I.
\eea
We can express the Hamiltonian in terms of their eigenvalues and bi-orthogonal eigenstates as
\bea
H(k)=\sum_n E_n(k) |u_{nR}(k)\ra\la u_{nL}(k)|.
\eea
Here, $\text{Im} E(k)\leq 0$ is satisfied because of the lossy nature of our models, and $\text{Im} E(k)=0$ is the imaginary gap closing point. For our purpose, it is convenient to start from the expression
\bea
E_n(k)=\frac{\la u_{nR}(k)|H(k)|u_{nR}(k)\ra}{\la u_{nR}(k)|u_{nR}(k)\ra}, \label{ERR}
\eea which is a simple consequence of $H(k)|u_{nR}(k)\ra= E_n(k)|u_{nR}(k)\ra$. Note that only right eigenstates appear here. The expression using both left and right eigenstates, $E_n(k)=\frac{\la u_{nL}(k)|H(k)|u_{nR}(k)\ra}{\la u_{nL}(k)|u_{nR}(k)\ra}$, though also valid, is less convenient for our subsequent calculations.

We now take the imaginary part of Eq. (\ref{ERR}),  which is \bea \text{Im} E_n(k)=\frac{1}{2i}[E_n(k)-E^*_n(k)] &=& \frac{1}{2i}[\frac{\la u_{nR}(k)|H(k)|u_{nR}(k)\ra}{\la u_{nR}(k)|u_{nR}(k)\ra} - \frac{\la u_{nR}(k)|H^\dag(k)|u_{nR}(k)\ra}{\la u_{nR}(k)|u_{nR}(k)\ra}] \nn\\ &=&\frac{\la u_{nR}(k)|D(k)|u_{nR}(k)\ra}{\la u_{nR}(k)|u_{nR}(k)\ra},
\eea
where we have defined \bea D(k)=[H(k)-H^{\dag}(k)]/2i, \eea which is a Hermitian matrix.  It can be diagonalized as
\bea
D(k)=\sum_i \lambda_i(k) |d_i(k)\ra\la d_i(k)|,
\eea with real-valued eigenvalues.
Inserting this formula into the expression of $\Im E(k)$, we see
\bea
\Im E(k)=\frac{\sum_i \lambda_i(k) |\la d_i(k)|u_{nR}(k)\ra|^2}{\la u_{nR}(k)|u_{nR}(k)\ra}.\label{relation1}
\eea
For the two models above, we observe that they both share a $k$-independent $D(k)=\frac{\gamma}{2}\sigma_z-\frac{\gamma}{2} I$. Its eigenvalues and corresponding eigenstates are
\bea
\lambda_1(k)=0,\ |d_1(k)\ra=(1,0)^T; \nn \\
\lambda_2(k)=-\gamma,\ |d_2(k)\ra=(0,1)^T.
\eea
Eq.~(\ref{relation1}) tells us that $\Im E(k)=0$ requires $|\la d_2(k)|u_{nR}(k)\ra|=0$, i.e. the right eigenstate at the imaginary gap closing point should be orthogonal to $|d_2(k)\ra=(0,1)^T$.  Thus, $|u_{nR}(k)\ra$ is parallel to $|d_1\ra$ or, equivalently, $|d_1\ra$ is a right eigenstate of $H(k)$ at the imaginary gap closing point. This is possible only when the coefficient of $\sigma_x$ in $H(k)$ vanishes. For our models, the $\sigma_x$ coefficient is $t_1+t_2 \cos k$, and therefore the imaginary gap closes at $k=k_0$ determined by \bea \cos k_0=-t_1/t_2.\eea  Since $|\cos(k)|\leq 1$ for real-valued $k$, we obtain the imaginary gap closing condition $|t_1|\leq |t_2|$. The energies of the imaginary gap closing points, where the $E(k)$ curve touches the real axis, are given by
\bea
\begin{aligned}
&\text{Model I}:\ \omega_0^\pm=\pm\sqrt{t_2^2-t_1^2},\\
&\text{Model II}:\omega_0^\pm=\frac{t_3}{t_2} \left(-t_1\cos \alpha\pm\sqrt{t_2^2-t_1^2}\sin \alpha\right)\ .
\end{aligned}
\eea

\section{ Explicit calculation of the bulk decay exponent }
In the main article, we have used the fact that the decay of bulk $P_x$ follows an algebraic law $P_x\sim |x-x_0|^{-\alpha_b}$ when the imaginary gap closes. Here, we provide an explicit calculation of the exponent $\alpha_b$. To this end, we consider a long one-dimensional chain so that the boundary effect is negligible, and denote $P_x$ by $P^\infty_x$, meaning that the chain is effectively infinite. We start from the formula
\bea
P_x^{\infty} =\frac{\gamma}{\pi}\int_{-\infty}^{+\infty} d\omega |\la x,B|G(\omega)|x_0,A\ra|^2 \label{Px}
\eea
with
\begin{align}
\la x,B|G(\omega)|x_0,A\ra&=\int_0^{2\pi}\frac{dk}{2\pi}  e^{ik(x-x_0)}\left(\frac{1}{\omega+i0^{+}-H(k)}\right)_{BA}\nn\\
                  &=\oint_{|\beta|=1} \frac{d\beta}{2\pi i\beta} \beta^{x-x_0} \left(\frac{1}{\omega+i0^{+}-H(\beta)}\right)_{BA}, \label{toeplitz}
\end{align}
where $H(\beta)\equiv H(k)|_{e^{ik}\rw\beta}$ \cite{yao2018edge,Xue2021simple}. For our specific model in the main article, \bea \left(\frac{1}{\omega+i0^{+}-H(\beta)}\right)_{BA}=\frac{t_1+t_2\frac{\beta+\beta^{-1}}{2}}{\det[\omega+i0^{+}-H(\beta)]}.\eea The integral in Eq. (\ref{toeplitz}) can be done by the residue theorem, and the asymptotic behavior at $|x-x_0|\rw\infty$ is determined by the roots of $\det[\omega+i0^{+}-H(\beta)]=0$ \cite{Xue2021simple}. As a quadratic equation, it has two roots $\beta_{L}(\omega)$ and $\beta_{R}(\omega)$, which satisfy $|\beta_{L}(\omega)|\geq 1\geq |\beta_{R}(\omega)|$ for real-valued $\omega$.  Following Ref. \cite{Xue2021simple}, we have $\la x,B|G(\omega)|x_0, A\ra\sim f_L\beta_L^{x-x_0}$ for $x<x_0$, and $\la x, B|G(\omega)|x_0,A\ra\sim f_R\beta_R^{x-x_0}$ for $x>x_0$, where the residue factors $f_{L/R}$ are $x$-independent and their values are
\bea
f_{L/R}(\omega)= \lim_{\beta\rightarrow \beta_{L/R}} (\beta-\beta_{L/R}) \frac{t_1+t_2\frac{\beta+\beta^{-1}}{2}}{\beta\det[\omega+i0^+ -H(\beta)]}.
\eea
Thus, Eq. (\ref{Px}) becomes
\bea
P_x^{\infty}=\frac{\gamma}{\pi}\int_{-\infty}^{+\infty}d\omega|f_{L/R}(\omega)|^2 |\beta_{L/R}(\omega)|^{2(x-x_0)} \label{bulkP},
\eea
where the subscript $L$ and $R$ corresponds to $x<x_0$ and $x>x_0$, respectively. As has been explained in the main article, the integral is dominated at large $|x-x_0|$ by the neighborhood of imaginary gap closing point $\omega_0$, where $|\beta_{L/R}(\omega_0)|=1$. To find the asymptotic behavior of the integral, we need the expansions of $\beta_{L/R}(\omega)$ and $f_{L/R}(\omega)$ near the imaginary gap closing point.

Hereafter, we shall focus on the $x<x_0$ region and therefore only $\beta_L(\omega)$ is relevant. Writing $\omega=\omega_0+\delta\omega$, and expanding $f_L(\omega),\beta_L(\omega)$ to the lowest order of $\delta\omega$, we have \bea |f_L(\omega)|^2\sim \delta\omega^m, \eea and $|\beta_L(\omega)|\approx 1+K\delta\omega^n\approx \exp(K\delta\omega^n)$ or, equivalently \bea \ln|\beta_L(\omega)|\approx K\delta\omega^n. \label{n} \eea  It follows that \bea P_x^{\infty} \sim \int d(\delta\omega)  \delta\omega^m \exp(-2K\delta\omega^n|x-x_0|) \sim |x-x_0|^{-(m+1)/n}, \eea and therefore $\alpha_b=(m+1)/n$. The rest part of this section is to calculate $n$ and $m$.

Let us calculate $n$ first. Note that the $i0^+$ term is irrelevant in most cases and we shall discard it for the moment, so that the determinant equation $\det[\omega+i0^{+}-H(\beta)]=0$ becomes $\det[\omega-H(\beta)]=0$.  For model I, namely the Eq. (1)(2) of the main article, the equation $\det[\omega-H(\beta)]=0$ can be explicitly written as
\bea
t_2 (t_1+\frac{\gamma}{2})\beta+t_2 (t_1-\frac{\gamma}{2})\beta^{-1}+t_1^2+t_2^2-\omega^2-i\gamma\omega=0.
\eea Its two roots are
\bea
\beta_{\pm}(\omega)=\frac{-b(\omega)\pm \sqrt{b^2(\omega)-4t_2^2(t_1^2-\gamma^2/4)}}{2t_2(t_1+\gamma/2)} \label{betapm}
\eea
where $b(\omega)=t_1^2+t_2^2-\omega^2-i\gamma\omega$. At the imaginary gap closing point of the Bloch Hamiltonian, $\omega_0^\pm=\pm\sqrt{t_2^2-t_1^2}$, there exists at least one root whose modulus is $1$.  To avoid excessive signs, we shall only focus on the positive frequency $\omega_0^{+}=\sqrt{t_2^2-t_1^2}$ (We focus on $t_2\geq t_1\geq0$ throughout this section). Under the ordering $|\beta_L(\omega)|\geq 1\geq |\beta_R(\omega)|$, the two roots read
\bea
\beta_L(\omega_0^+)=\beta_{-}(\omega_0^+)=-\frac{t_1}{t_2}+i\frac{\sqrt{t_2^2-t_1^2}}{t_2},\ \beta_R(\omega_0^+)=\beta_{+}(\omega_0^+)=\frac{t_1-\gamma/2}{t_1+\gamma/2}\left(-\frac{t_1}{t_2}-i\frac{\sqrt{t_2^2-t_1^2}}{t_2}\right).
\eea
We can rewrite it as
\bea
\beta_{L}(\omega_0^+)=e^{ik_0},\ \beta_{R}(\omega_0^+)=\frac{t_1-\gamma/2}{t_1+\gamma/2} e^{-ik_0}
\eea
where $k_0=\arccos(-t_1/t_2)$. As the frequency slightly shifts away from the imaginary gap closing point $\omega_0^+$, namely $\omega=\omega_0^{+}+\delta \omega$,  $\beta_{L}(\omega_0^{+}+\delta \omega)$ will also shift from $\exp(ik_0)$ to
$\exp[i(k_0+\delta k+i\delta k')]$, $\delta k$ and $\delta k'$ being functions of $\delta\omega$. The leading order expansion of $i(\delta k+i\delta k')=\ln[\beta_L(\omega)/\beta_L(\omega_0)]$ is
\bea
i(\delta k+i\delta k')=\left(\frac{\partial \ln \beta_L}{\partial \omega}\right)_{\omega=\omega_0^{+}}\delta \omega+\left(\frac{\partial^2 \ln \beta_L}{\partial \omega^2}\right)_{\omega=\omega_0^{+}}\frac{\delta \omega^{2}}{2}+\cdots. \label{expansion}
\eea
The real part of left hand side (LHS) is $-\delta k'=\ln|\beta_L|\approx K\delta\omega^n$, and that of the right hand side (RHS) should tell us the value of $n$ of Eq. (\ref{n}). In fact, the lowest order nonzero real-valued coefficient of RHS should be identified as the $K\delta\omega^n$ term.
By explicit calculations, we obtain the derivatives
\bea
\begin{aligned}
&\left(\frac{\partial \ln \beta_L}{\partial \omega}\right)_{\omega=\omega_0^{+}}=\left(\frac{1}{\beta_L}\frac{\partial \beta_L}{\partial \omega}\right)_{\omega=\omega_0^{+}}=-\frac{i}{t_1},\\
&\left(\frac{\partial^2 \ln \beta_L}{\partial \omega^2}\right)_{\omega=\omega_0^{+}}= \left(\frac{1}{\beta_L}\frac{\partial^2 \beta_L}{\partial \omega^2}\right)_{\omega=\omega_0^{+}}-\left(\frac{1}{\beta_L}\frac{\partial \beta_L}{\partial \omega}\right)^2_{\omega=\omega_0^{+}}=\frac{\gamma\sqrt{t_2^2-t_1^2}}{t_1^3(2\omega_0^{+}+i\gamma)}.
\end{aligned}
\eea When $t_1=0$, the derivatives diverge and the expansion fails; let us focus on the $t_1\in(0,t_2]$ region for now and come back to the special case $t_1=0$ shortly. If $t_1\in(0,t_2)$, the leading order real-valued term at the RHS of Eq. (\ref{expansion}) is the $\delta \omega^2$ term; therefore, we have $-\delta k'=\ln|\beta_L| \propto \delta \omega^2$, meaning that $n=2$. On the other hand,  if $t_1=t_2$,  the second-order term at RHS vanishes and higher order expansion is necessary. The lowest order contribution occurs at the fourth order of $\delta \omega$:
\bea
t_1=t_2:\ \left(\frac{\partial \ln \beta_L}{\partial \omega}\right)_{\omega=\omega_0^{+}}=-\frac{i}{t_1},\ \left(\frac{\partial^2 \ln \beta_L}{\partial \omega^2}\right)_{\omega=\omega_0^{+}}=0,\ \left(\frac{\partial^3 \ln \beta_L}{\partial \omega^3}\right)_{\omega=\omega_0^{+}}=-\frac{i}{t_1^3},\ \left(\frac{\partial^4 \ln \beta_L}{\partial \omega^4}\right)_{\omega=\omega_0^{+}}=\frac{6}{\gamma t_1^3},
\eea and we have $n=4$.

To obtain the value of $m$, we calculate the residue factor $f_L(\omega)$, which reads
\bea
f_L(\omega)= \lim_{\beta\rightarrow \beta_L} (\beta-\beta_L) \frac{t_1+t_2\frac{\beta+\beta^{-1}}{2}}{\beta\det[\omega-H(\beta)]}.
\eea
The determinant in the denominator, $\det[\omega-H(\beta)]$, can be expressed in terms of its two roots:
\bea
\det[\omega-H(\beta)]=t_2(t_1+\gamma/2)(\beta-\beta_L)(\beta-\beta_R)/\beta.
\eea
Thus, in terms of the roots, $f_L(\omega)$ reads
\bea
f_L(\omega)= \frac{t_1+t_2\frac{\beta_L+\beta_L^{-1}}{2}}{t_2(t_1+\gamma/2)(\beta_L-\beta_R)}.\label{residue2}
\eea
Making use of the results from the previous section, we have $f_L(\omega)=0$ at the imaginary gap closing point, which follows from
\bea
t_1+t_2\frac{\beta_L(\omega_0^{+})+\beta_L^{-1}(\omega_0^{+})}{2}=t_1+t_2 \cos k_0=0.
\eea
Near the gap closing point, the leading order expansion of $f_L(\omega)$ can be obtained by using the aforementioned expressions of $\beta_L$ and $\beta_R$. We have
\bea
f_L(\omega_0^++\delta \omega)\approx -\frac{\sin k_0}{\gamma\cos k_0+2i t_1\sin k_0}(\delta k+i\delta k')
\eea
when $t_1\neq t_2$ and
\bea
f_L(\omega_0^++\delta \omega)\approx -\frac{(\delta k+i\delta k')^2}{2\gamma}
\eea
when $t_1=t_2$ and $k_0=\pi$. It follows that $|f_L(\omega_0^++\delta \omega)|^2\propto |\delta k+i\delta k'|^2\propto \delta\omega^2$ when $0<t_1<t_2$, and $|f_L(\omega_0^++\delta \omega)|^2\propto |\delta k+i\delta k'|^4\propto \delta\omega^4$ when $t_1=t_2$. In other words, we have $m=2$ for $t_1\in(0,t_2)$, and $m=4$ for $t_1=t_2$.

Now we come back to the special case $t_1=0$, for which the Taylor expansion fails due to divergent derivatives. Therefore, we follow  a different approach. We insert the frequency $\omega=\omega_0^+ +\delta \omega=t_2+\delta \omega$ into the expression of $\beta_\pm$ [Eq. (\ref{betapm})], which yields
\bea
\beta_\pm(t_2+\delta \omega)\approx i\pm2\sqrt{\frac{i(t_2+i\gamma/2)\delta \omega}{
\gamma t_2}},
\eea
and consequently,
\bea
\ln |\beta_\pm(t_2+\delta \omega)|\propto \sqrt{|\delta \omega|}.
\eea
This gives a fractional value $n=1/2$ for $t_1=0$.  Regarding $f_L(\omega)$, we observe that $\beta_+( \omega_0^+ )= \beta_- (\omega_0^+)$ or, equivalently, $\beta_L( \omega_0^+ )= \beta_R (\omega_0^+)$ (recall that $\beta_{L/R}$ are just $\beta_\pm$ under the ordering $|\beta_{L}|\geq |\beta_{R}|$), at the imaginary gap closing point, so that Eq.~(\ref{residue2}) becomes ill-defined.
To cope with this issue, we have to restore the infinitesimal $i0^+$ term in the determinant equation $\det[\omega+i0^{+}-H(\beta)]=0$, namely,
\bea
f_L(\omega)&=&\lim_{\epsilon\rightarrow 0^+} \lim_{\beta\rightarrow \beta_L} (\beta-\beta_L) \frac{t_1+t_2\frac{\beta+\beta^{-1}}{2}}{\beta\det[\omega+i\epsilon-H(\beta)]} \nn \\
&=& \lim_{\epsilon\rightarrow 0^+} \frac{\beta_L(t_2+i\epsilon)+\beta_L^{-1}(t_2+i\epsilon)}{\beta_L(t_2+i\epsilon)-\beta_R(t_2+i\epsilon)}\frac{1}{\gamma} \nn\\
&=& \frac{1}{\gamma}
\eea
where $\beta_{L/R}(t_2+i\epsilon)=i\left(1\pm2\sqrt{(t_2+i\gamma/2)\epsilon/
\gamma t_2}\right)$. In the second line, we observe that the numerator $\beta_L(t_2+i\epsilon)+\beta_L^{-1}(t_2+i\epsilon)\propto \ \sqrt{\epsilon}$ and the denominator $\beta_L(t_2+i\epsilon)-\beta_R(t_2+i\epsilon)\propto \ \sqrt{\epsilon}$ in the same way, which causes their ratio $\frac{\beta_L(t_2+i\epsilon)+\beta_L^{-1}(t_2+i\epsilon)}{\beta_L(t_2+i\epsilon)-\beta_R(t_2+i\epsilon)}\rw 1$ as $\epsilon\rw 0$.  The nonzero value of $f_L$ at the imaginary gap closing point means that $m=0$ for $t_1=0$.

\begin{figure}
\includegraphics[width=4.4cm, height=3.9cm]{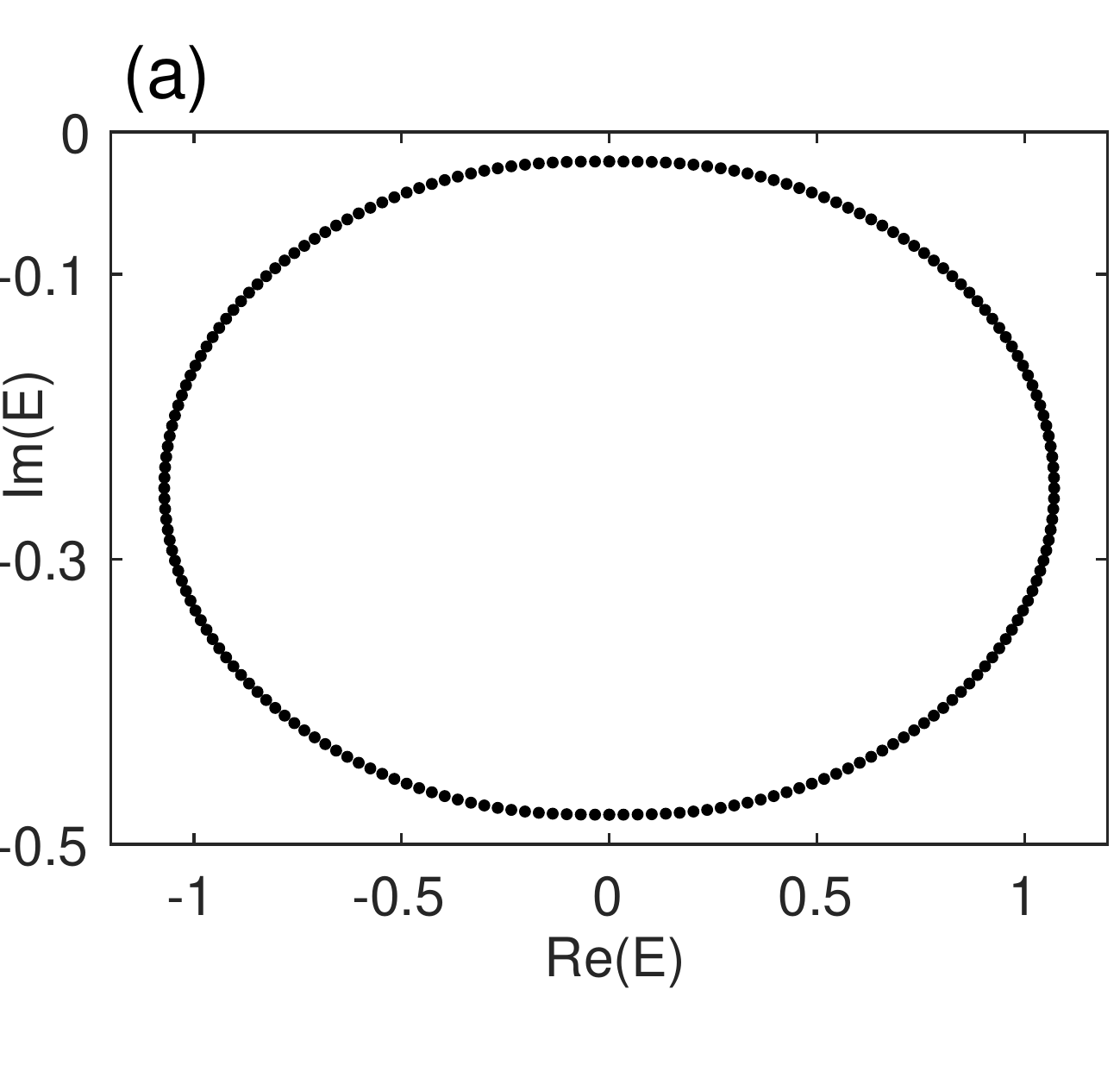}
\includegraphics[width=4.4cm, height=3.9cm]{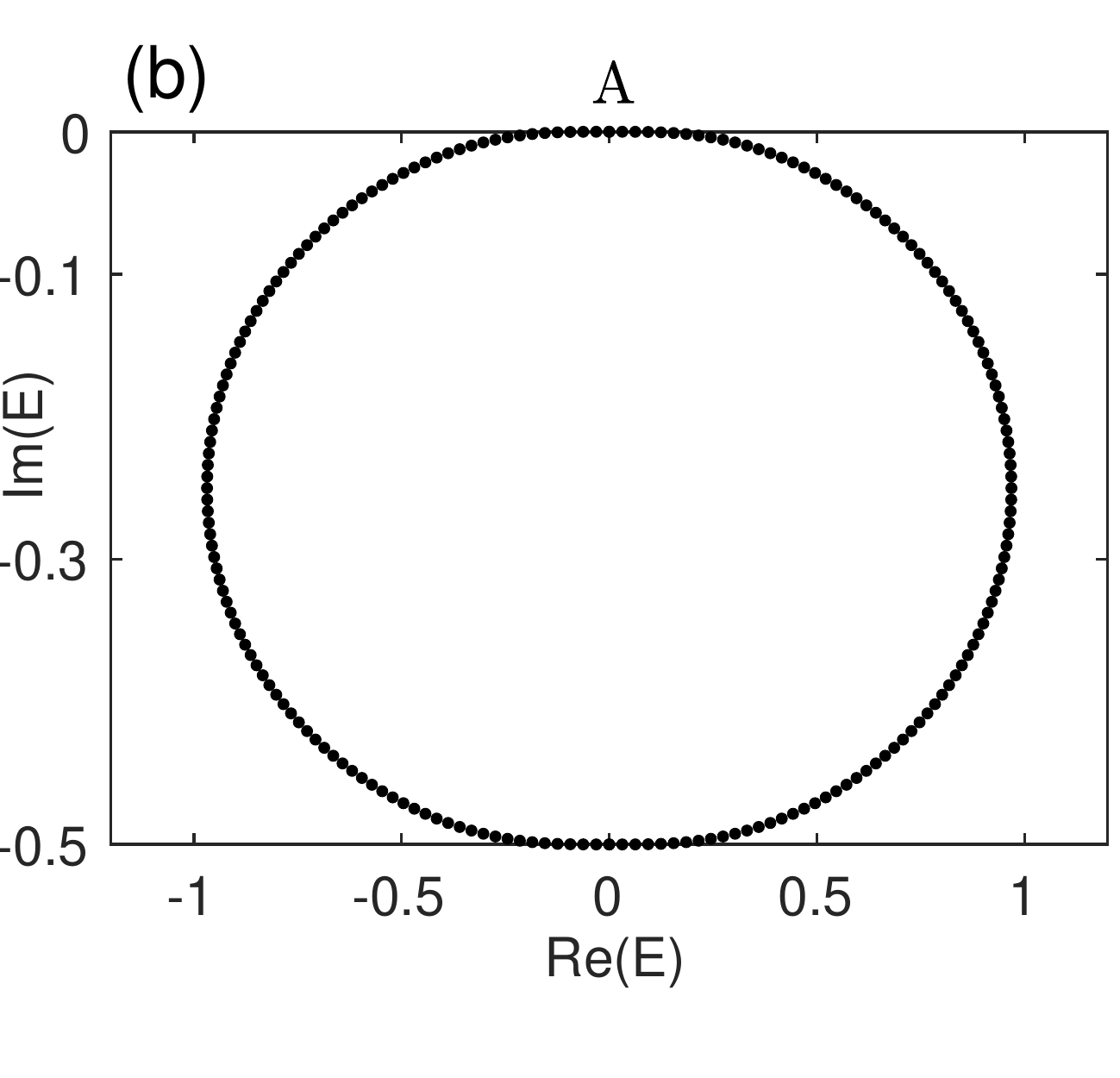}
\includegraphics[width=4.4cm, height=3.9cm]{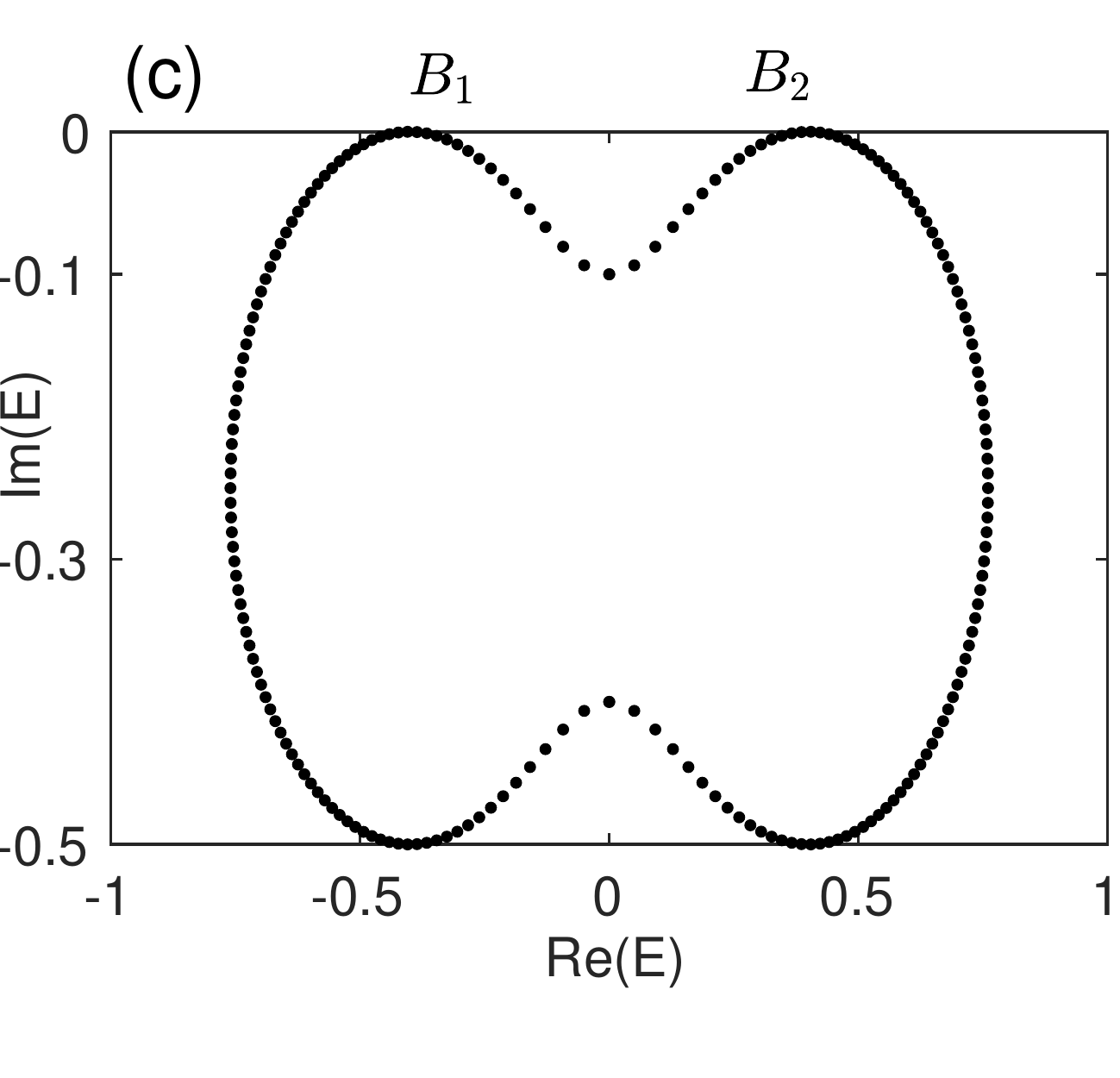}
\includegraphics[width=4.4cm, height=3.9cm]{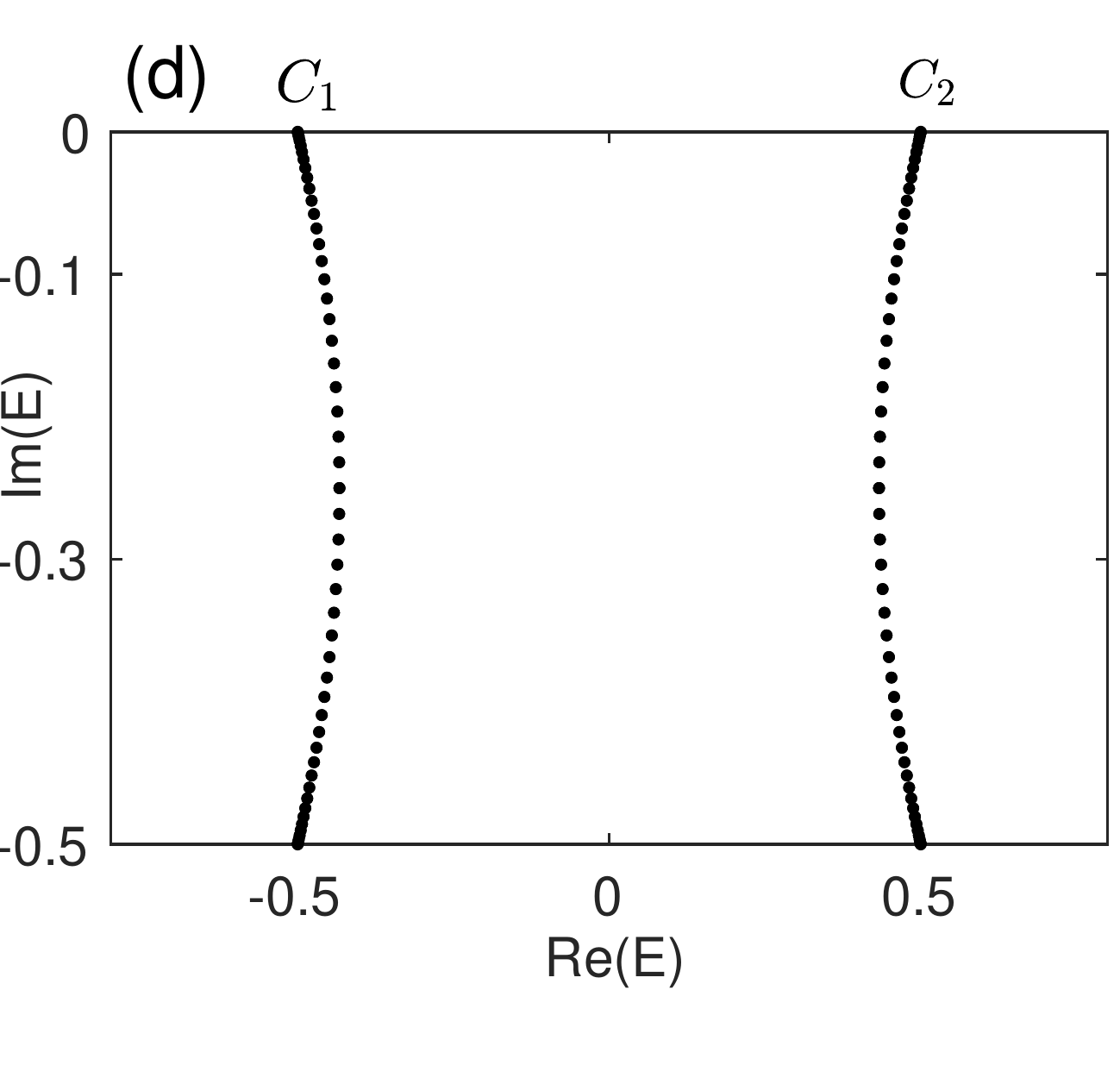}
\includegraphics[width=4.4cm, height=3.9cm]{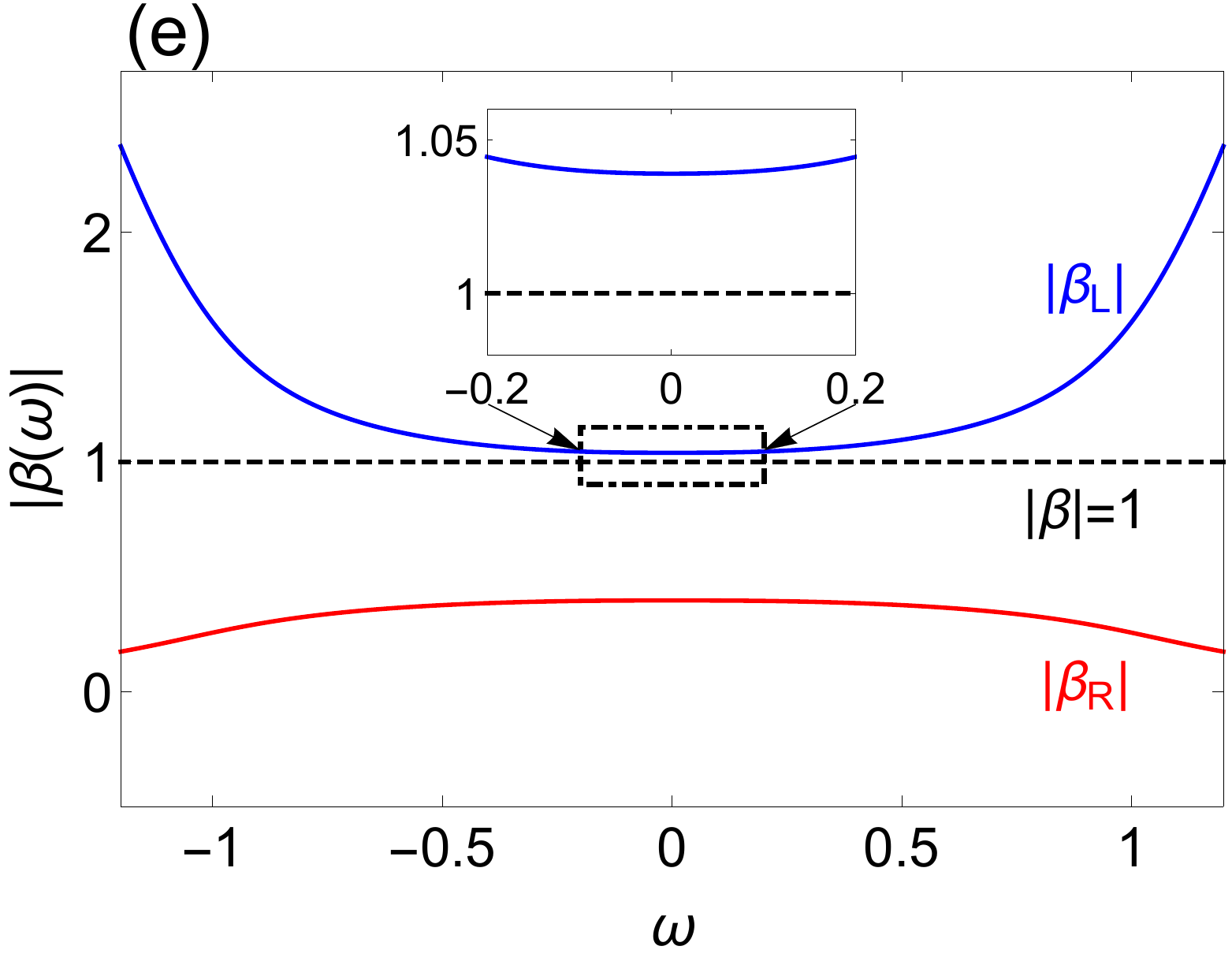}
\includegraphics[width=4.4cm, height=3.9cm]{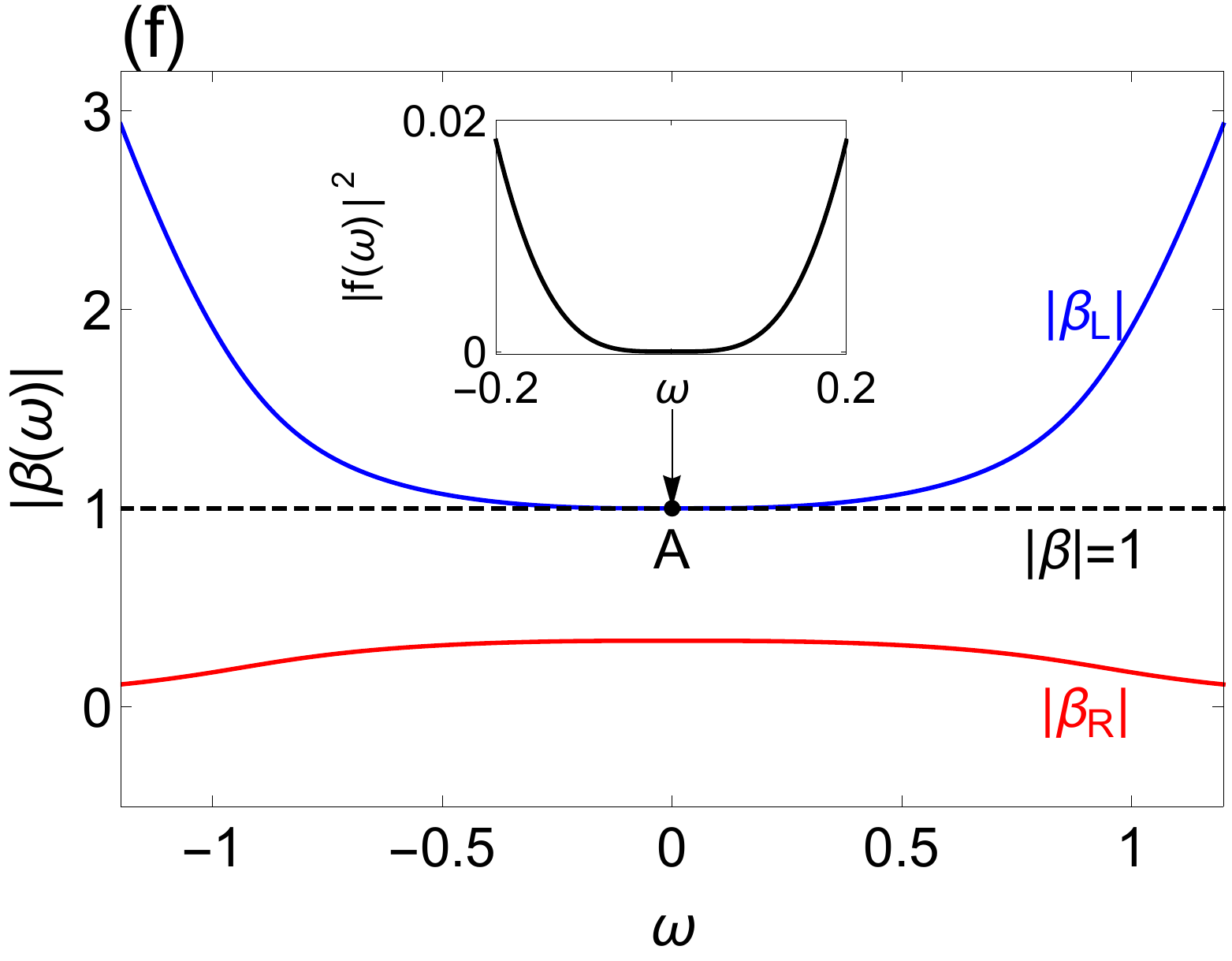}
\includegraphics[width=4.4cm, height=3.9cm]{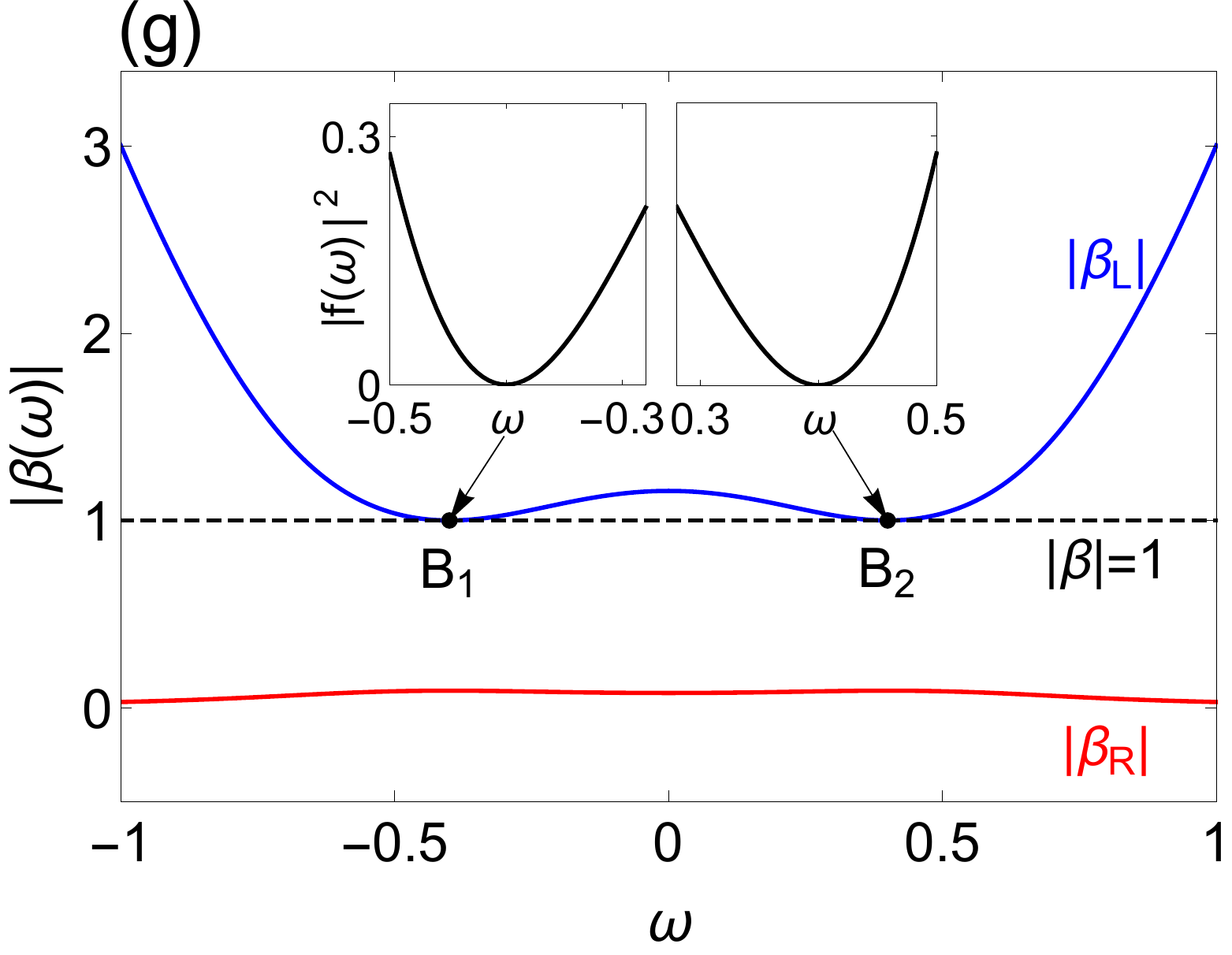}
\includegraphics[width=4.4cm, height=3.9cm]{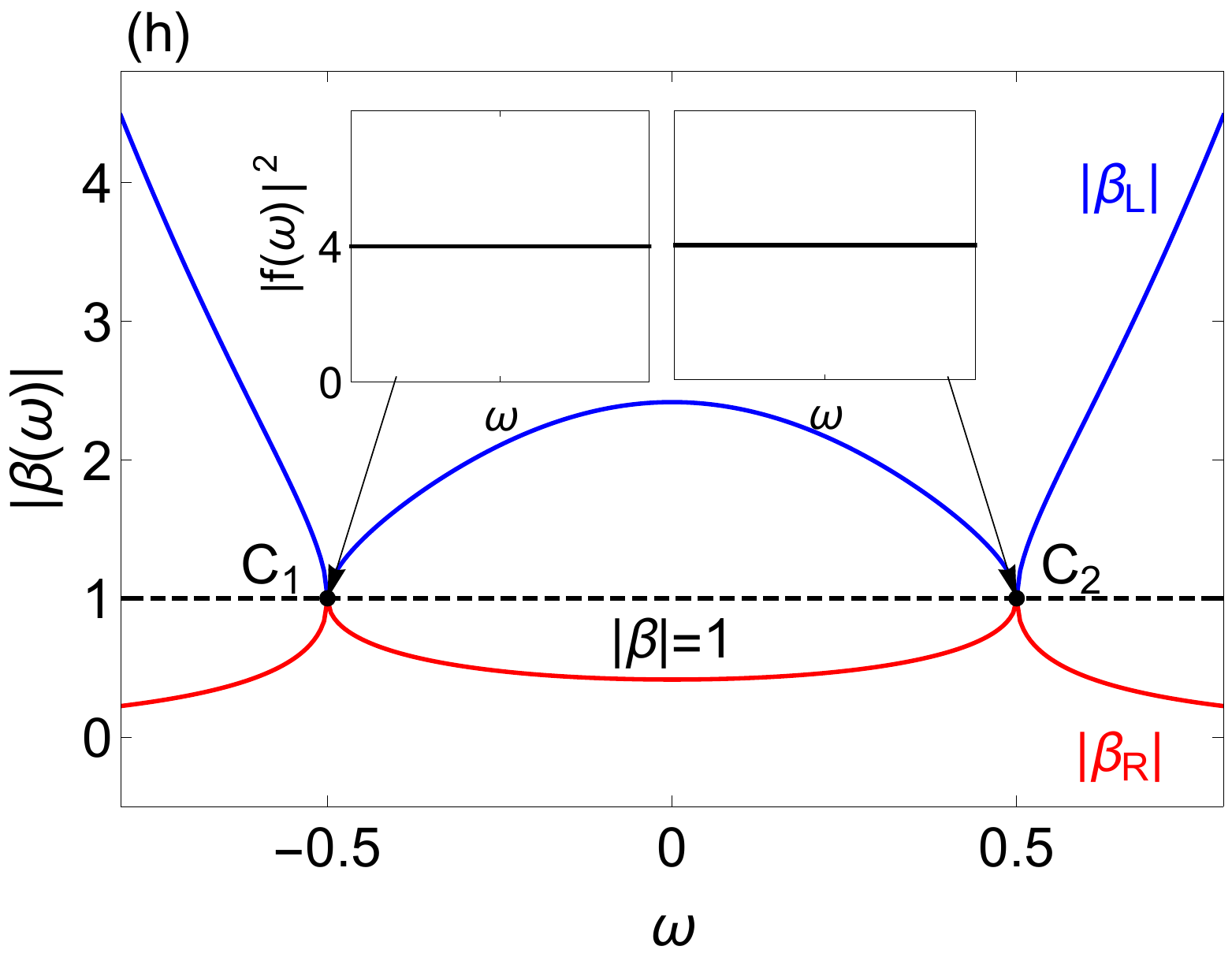}
\caption{The Bloch energy spectrums (upper panels) versus $|\beta(\omega)|$ (lower panel) of model I. The parameters are $t_2=0.5, \gamma=0.5$ and  $t_1=0.6>t_2$ in (a)(e), $t_1=0.5=t_2$ in (b)(f), $t_1=0.3<t_2$ in (c)(g), and $t_1=0$ in (d)(h). The inset in (e) shows the zoom-in gap between $|\beta_L|$ and $|\beta|=1$, and the insets in (f)(g)(h) show the frequency dependence of the residue near the imaginary gap closing points. }
\label{S1}
\end{figure}

The main results of this section can be summarized as
\bea
\begin{aligned}
&t_1=0:\ \qquad \ln |\beta_L(\omega_0+\delta \omega)| \propto  \sqrt {|\delta\omega|},\ |f_L(\omega)|^2\propto \delta \omega ^0, \, (n,m)=(\frac{1}{2},0);\\
&t_1\in (0,t_2):\ \ln |\beta_L(\omega_0+\delta \omega)| \propto \delta \omega ^2,\ \quad |f_L(\omega)|^2\propto \delta \omega ^2,\, (n,m)=(2,2);\\
&t_1=t_2:\qquad \ln |\beta_L(\omega_0+\delta \omega)| \propto \delta \omega ^4,\ \quad |f_L(\omega)|^2\propto \delta \omega ^4, \, (n,m)=(4,4) \\
\end{aligned}
\eea
where $n,m$ stand for the exponents in the expansions $|\beta_L(\omega)|\approx  1+K\delta\omega^n$, and $|f(\omega)|^2\sim \delta\omega^m$. These analytic results are also confirmed numerically; see  Fig.~\ref{S1} of this Supplemental Material.

Inserting these expressions back into Eq.~(\ref{bulkP}), we find that (for $x<x_0$) the large $|x-x_0|$ behavior is $P_x^{\infty} \sim \int d(\delta\omega)  \delta\omega^m \exp(-2K\delta\omega^n|x-x_0|) \sim |x-x_0|^{-(m+1)/n}$ when $t_1\leq t_2$. In other words, we have $P_x^\infty\sim |x-x_0|^{-\alpha_b}$  with exponent $\alpha_b=(m+1)/n$ whose explicit value is
\bea
\begin{aligned}
t_1=0:\ \alpha_b=2;\\
t_1\in (0,t_2):\ \alpha_b=\frac{3}{2};\\
t_1=t_2:\ \alpha_b=\frac{5}{4}.\\
\end{aligned}
\eea

\section{Green's function formulas satisfy the sum rule}

We denote the onsite loss probability by $\gamma_j$ for each site $j=(x,A/B)$; for our specific model, $\gamma_{x,A}=0$ and $\gamma_{x,B}=\gamma>0$. The wavefunction norm evolves as
\bea
\frac{d}{dt}\la\psi(t)|\psi(t)\ra=-2\sum_j \gamma_j|\la j|\psi(t)\ra|^2<0.
\eea
Integration of both the left hand side (LHS) and right hand side (RHS) of this equation from $t=0$ to $t=\infty$ leads to
\bea
\la\psi(0)|\psi(0)\ra-\la\psi(\infty)|\psi(\infty)\ra = \sum_j P_j, \label{Pjsum}
\eea
where $P_j=2\gamma_j\int_0^\infty  dt   |\la j|\psi(t)\ra|^2$ is the local loss. Under the standard normalization $\la\psi(0)|\psi(0)\ra=1$ and $\la\psi(\infty)|\psi(\infty)\ra=0$, Eq. (\ref{Pjsum}) indeed gives $\sum_j P_j=1$.

We would like to do a consistency check that the sum rule $\sum_j P_j=1$ is satisfied in our Green's function approach, in which
\bea
P_j=\frac{\gamma_j}{\pi}\int_{-\infty}^{+\infty} d\omega |\la j|G(\omega)|\psi(0)\ra|^2 = \frac{\gamma_j}{\pi}\int_{-\infty}^{+\infty} d\omega |\la j|\frac{1} {\omega+i0^{+}-H}|\psi(0)\ra|^2.
\eea
Now the sum is
\bea
\begin{aligned}
\sum_j P_j &=\frac{1}{\pi}\int_{-\infty}^{+\infty} d\omega \sum_j\gamma_j\la\psi_0|G^\dag(\omega) |j\ra\la j|G(\omega)|\psi(0)\ra\\
&=\frac{1}{2\pi i}\int_{-\infty}^{+\infty}d\omega\la \psi(0)|G^\dag(\omega)(H^\dag-H)G(\omega)|\psi(0)\ra\nn
\\ &=\frac{1}{2\pi i}\int_{-\infty}^{+\infty}d\omega\la \psi(0)|G^\dag(\omega)[(\omega-H) -(\omega-H^\dag)]G(\omega)|\psi(0)\ra\nn \\
&=\frac{1}{2\pi i}\int_{-\infty}^{+\infty}d\omega\la \psi(0)|[G^\dag(\omega)-G(\omega)]|\psi(0)\ra\nn\\
  &=\sum_n\la \psi(0)|nR\ra\la nL| \psi(0)\ra\nn\\
  &=1,
\end{aligned}
\eea
where we have used $\sum_j\gamma_j|j\ra\la j|=(H^\dag-H)/2i$, and the spectral representation $G(\omega)=\sum_n |nR\ra\la nL|/(\omega+i0^+-E_n)$ in terms of the eigenstates $|nR\ra, |nL\ra$ and eigen-energies $E_n$. Note that the integrand \bea G^\dag(\omega)-G(\omega)\rw 1/\omega^2 \eea as $\omega\rw\infty$, and therefore the $\omega$ integration is well defined; taking contour in the upper or lower complex plane yields the same result.

The simple sum rule $\sum_j P_j=1$ places a constraint that $\alpha_b>1$ in the algebraic decay $P_x\sim |x-x_0|^{-\alpha_b}$, otherwise the integral
\bea
 \int_{-\infty}^{0} |x-x_0|^{-\alpha_b}  dx
\eea is divergent and the sum rule cannot be satisfied in an infinite chain.

\section{Bipolar non-Hermitian skin effect and bipolar edge burst}

\begin{figure}
\includegraphics[width=4.2cm, height=4cm]{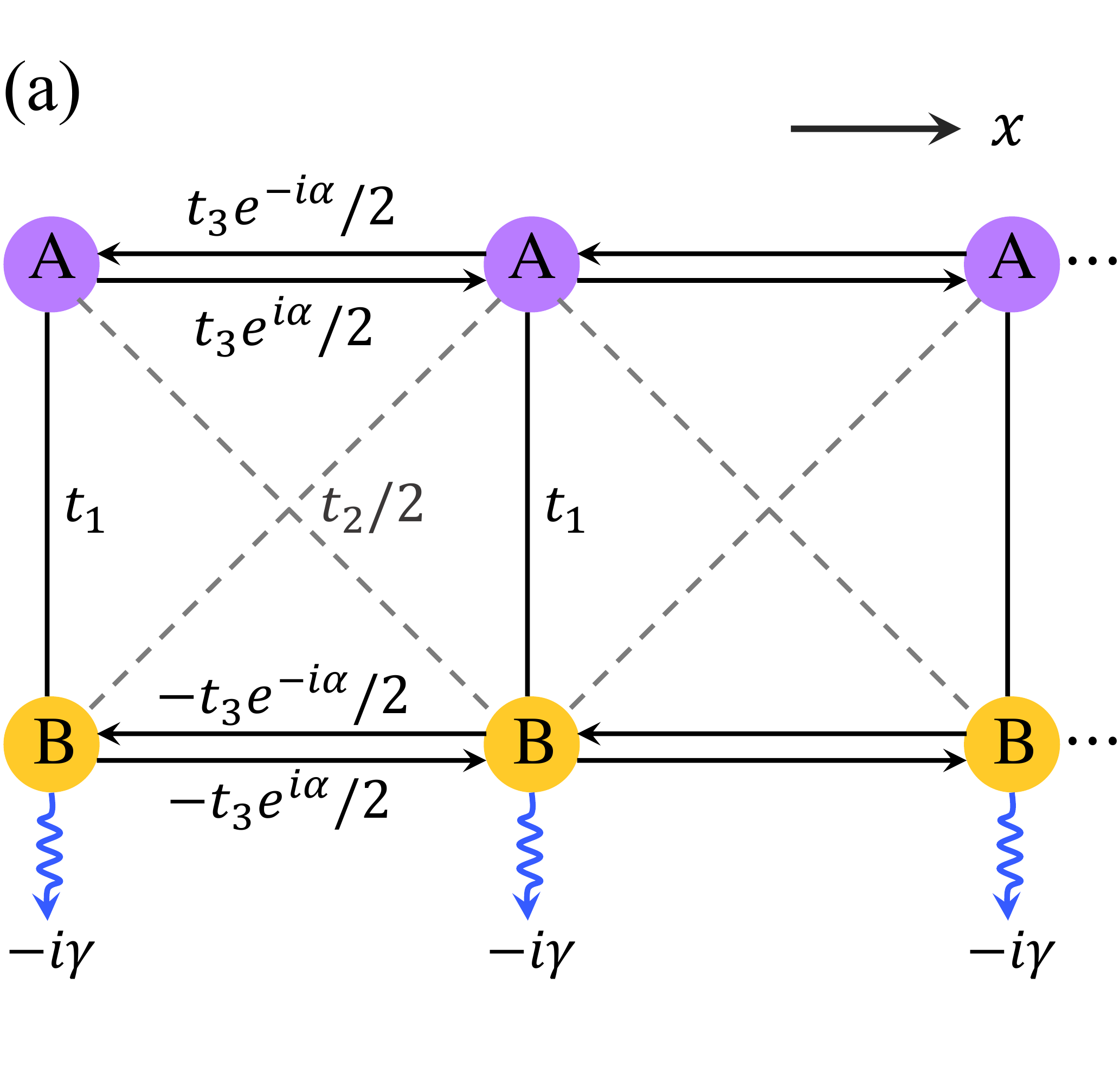}
\includegraphics[width=4.2cm, height=4cm]{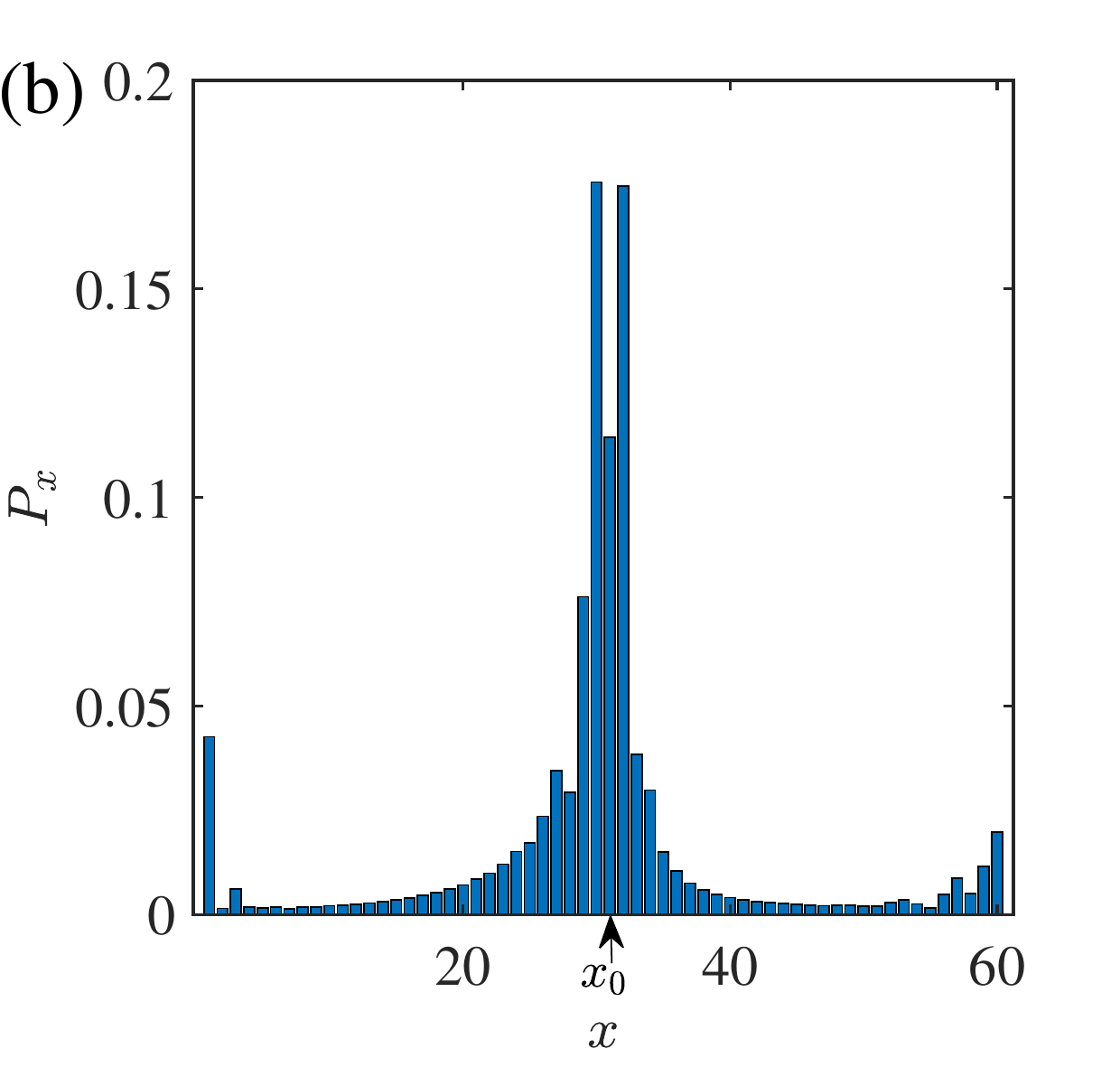}
\includegraphics[width=4.2cm, height=4cm]{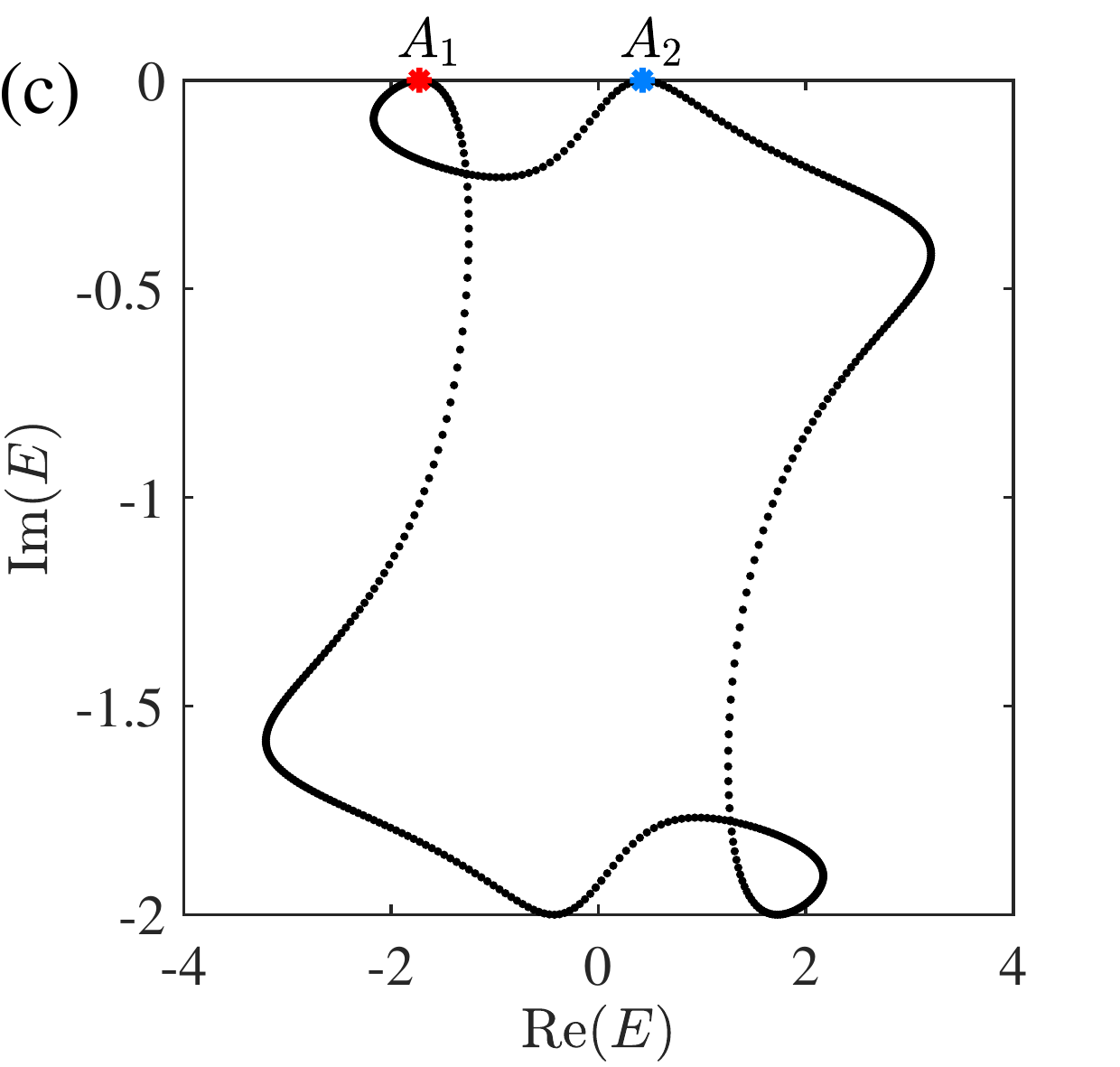}
\includegraphics[width=4.2cm, height=4cm]{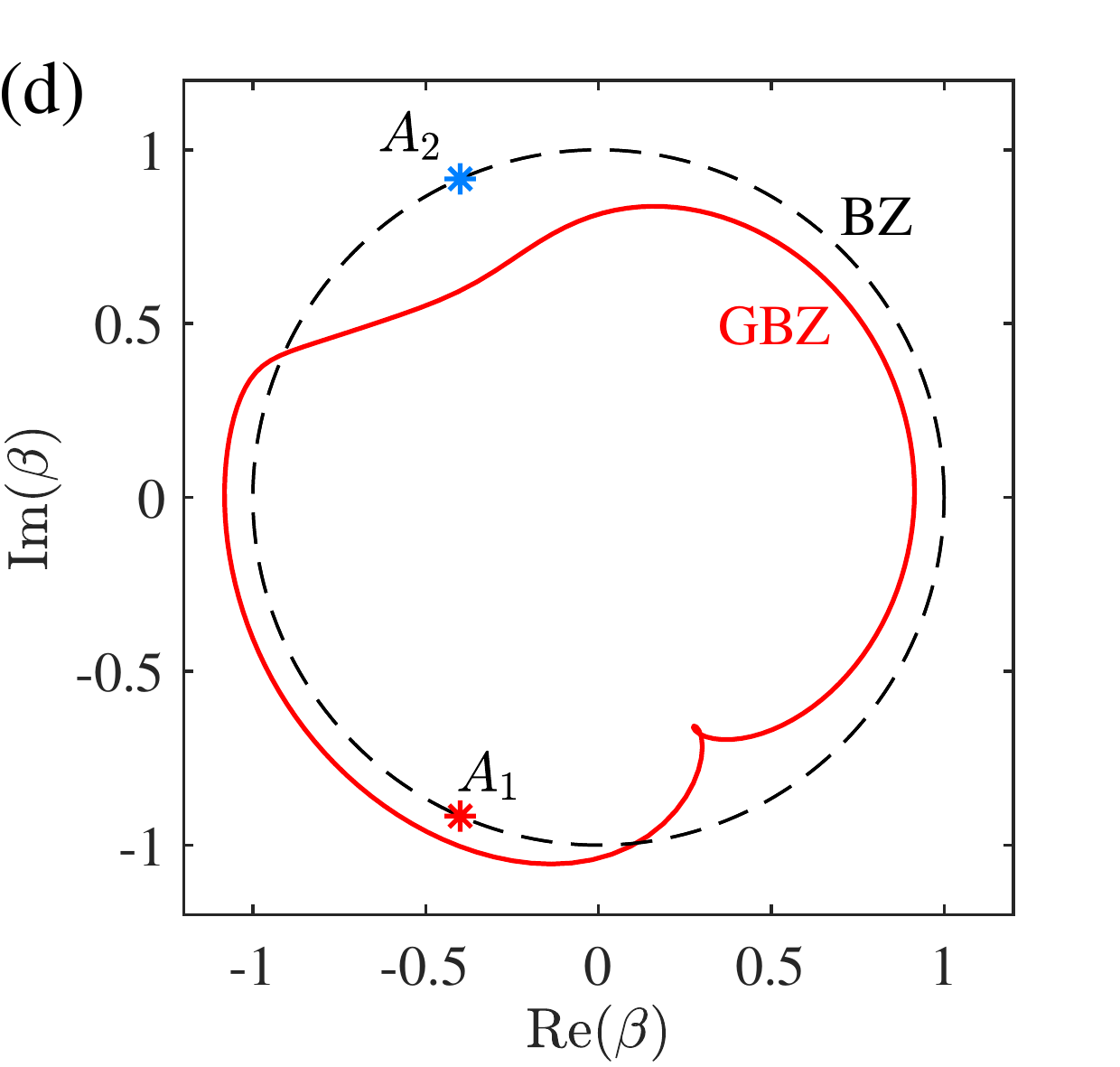}
\caption{(a) Pictorial illustration of the Hamiltonian with bipolar NHSE (Eq.~\ref{model2}). (b) $P_x$ profile of a finite chain with $L=60$. $x_0=31$. (c) The PBC spectrum. (d) The GBZ.  $t_1=0.8,t_2=2,t_3=2,\alpha=\pi/5;\gamma=2$. The imaginary gap closing points $A_1$ and $A_2$ in (c) correspond to $\beta$ value $A_1$ and $A_2$ in (d), respectively. The point enclosed by the GBZ ($A_1$) corresponds to edge burst at the right end, while that outside the GBZ ($A_2$) corresponds to edge burst at the left end.  }
\label{fig4}
\end{figure}

Based on our present understanding of the edge burst, we may expect that the bipolar non-Hermitian skin effect (NHSE) \cite{Song2019real}, namely, skin modes localized at both the left and right ends, might generate edge burst at both ends provided that the imaginary gap closes.  We have confirmed this expectation using the model II, whose Hamiltonian is reproduced as:
\begin{align}
H(k)=(t_1+t_2\cos k) \sigma_x+[t_3\cos(k-\alpha)+\frac{i\gamma}{2}]\sigma_z-\frac{i\gamma}{2}I, \label{model2}
\end{align}
and also illustrated in Fig.~\ref{fig4}(a) of this Supplemental Material. As shown in Fig.~\ref{fig4}(b), we have bipolar edge burst, namely, edge burst occurs at both ends of the chain.

Based on the bulk-edge scaling relation underlying the edge burst, we infer algebraic decay of $P_x$ in the bulk towards both left and right ends. Notably, one of the imaginary gap closing point ($A_1$) is enclosed by the GBZ, and the other ($A_2$) is outside the GBZ [Fig.~\ref{fig4}(d)]. This is in line with the GBZ-based formulas of Green's function, in which leftward and rightward Green's function is determined by roots outside and inside the GBZ \cite{Xue2021simple}.

\section{Random starting point}

\begin{figure}
\includegraphics[width=8cm, height=4cm]{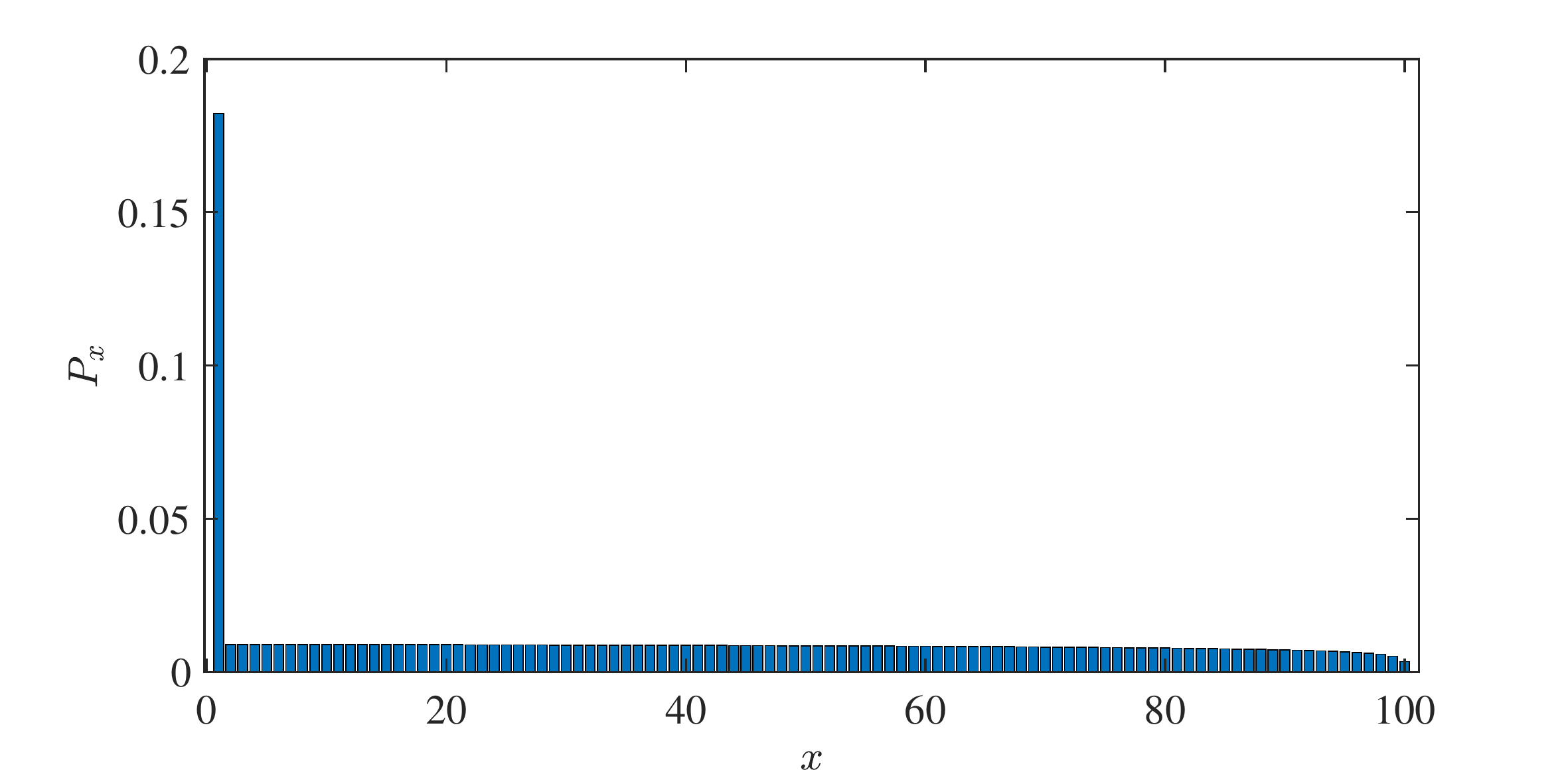}
\caption{The spatially resolved loss $P_x$ for a uniform distribution of starting point on an open-boundary chain. The model is the one from the main article (i.e. the model I of this Supplemental Material). The parameters are $t_1=0.4, t_2=0.5, \gamma=0.8$. We take chain length $L=100$, and consequently $p_s=1/L=1/100$.  }
\label{S2}
\end{figure}

In the main article, the walker starts from a fixed location $x_0$. Suppose that the initial location is randomly distributed in the chain of length $L$, then we have
\bea
P_x=\sum_{s} P(s\rightarrow x) p_{s},
\eea
where $p_s$ is the probability of starting from location $s$, with the sum rule $\sum_s p_s=1$ satisfied. In the main article, we have focused on the case $p_{s}=\delta_{s,x_0}$. Another interesting choice of $p_s$ is the uniform distribution $p_s=1/L$ that is location-independent. Such a distribution describes an incoherent input that has equal probability at everywhere along the chain. We find that the edge burst remains present for a generic $p_s$ distribution. Specifically, we have shown $P_x$ for the uniform $p_s$ distribution in Fig.~\ref{S2} of this Supplemental Material, with a prominent edge burst. The relative height of the edge burst can be readily estimated. Because the typical distance between the starting point and the edge is of order $L$, the height of edge peak is roughly $P_\text{edge}\sim L^{-\alpha_e}\sim L^{-\alpha_b+1}$. On the other hand, the average height of $P_x$ in the bulk is $L^{-1}$. Therefore, the relative height of the edge peak is roughly $L^{-\alpha_b+1}/L^{-1}\sim L^{-\alpha_b+2}$. For example, with $\alpha_b=3/2$, the relative height for $t_1\in(0,t_2)$ is $\sqrt{L}$, which grows with $L$.

\section{Hamiltonian with k-dependent anti-Hermitian part}

\begin{figure}
\includegraphics[width=8cm, height=3cm]{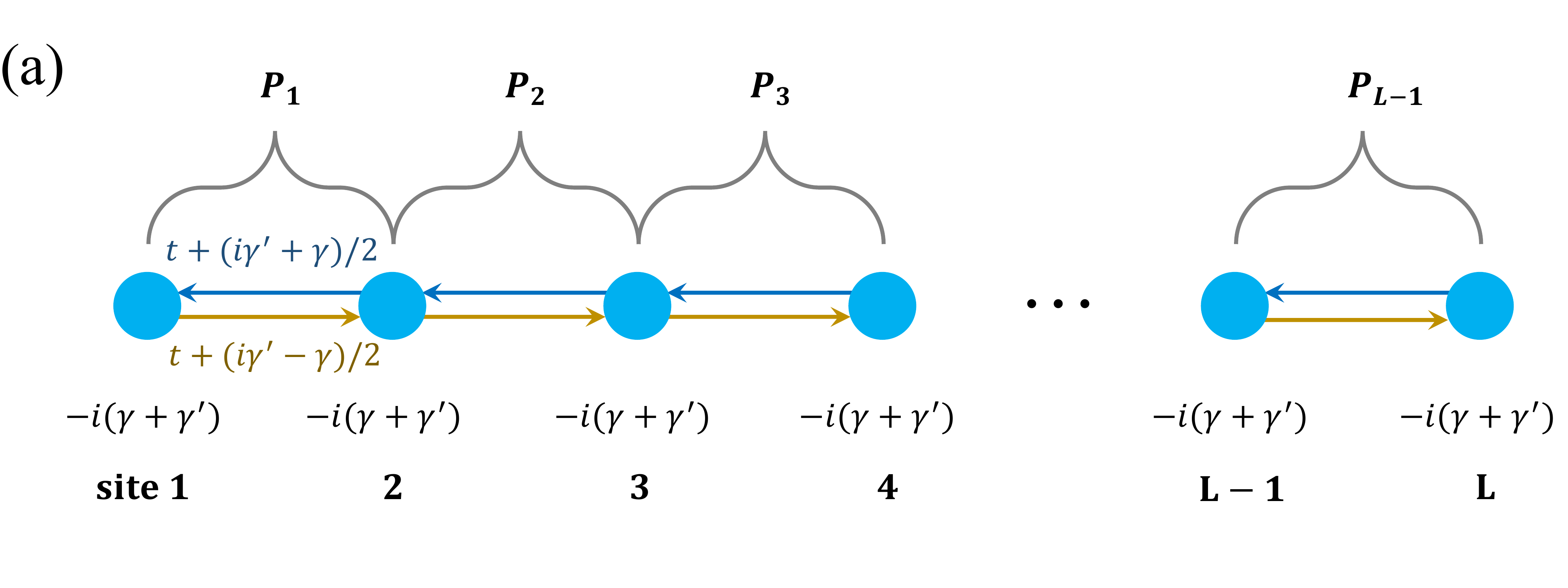}
\includegraphics[width=8cm, height=3cm]{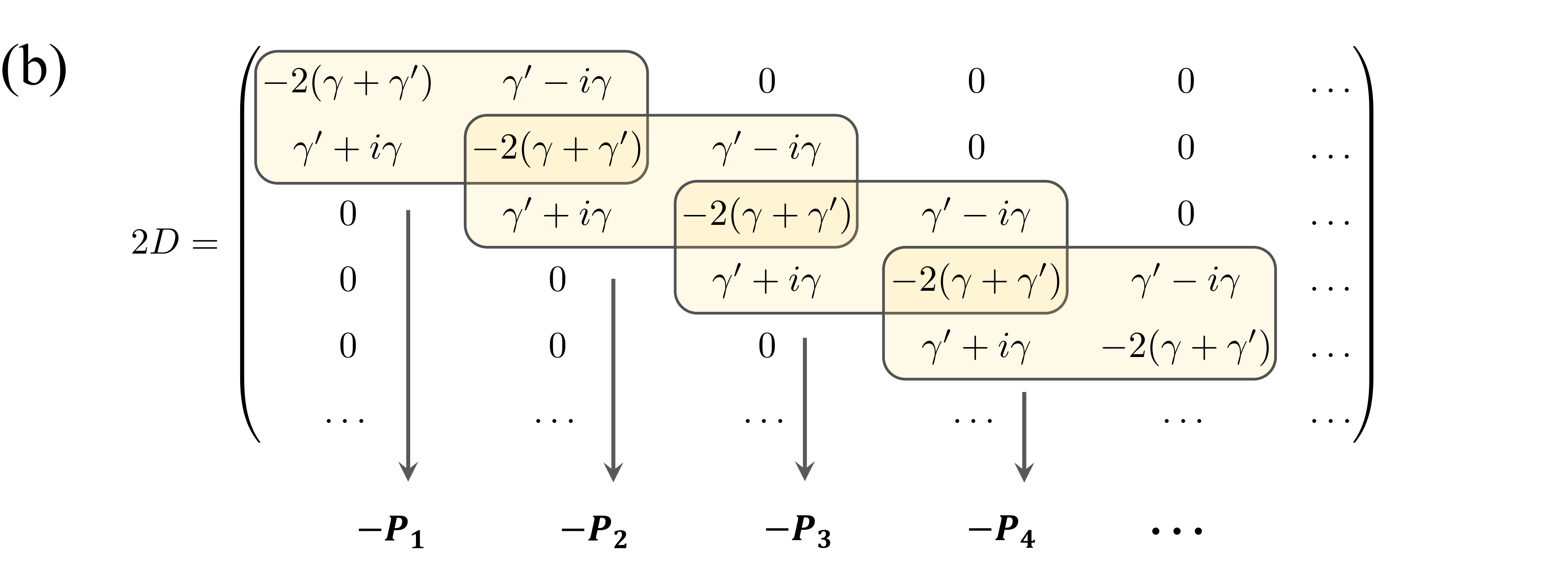}
\includegraphics[width=8cm, height=4cm]{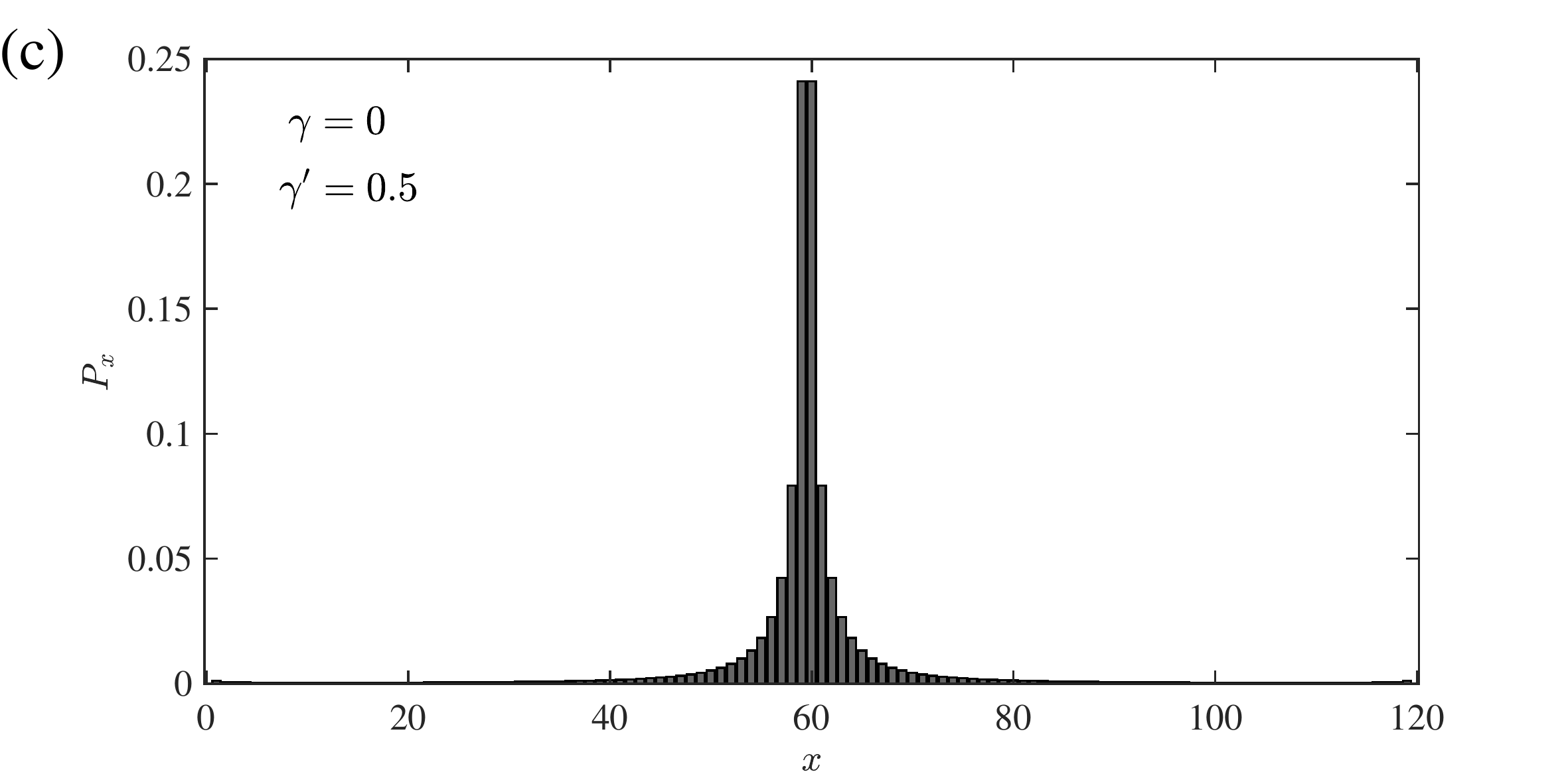}
\includegraphics[width=8cm, height=4cm]{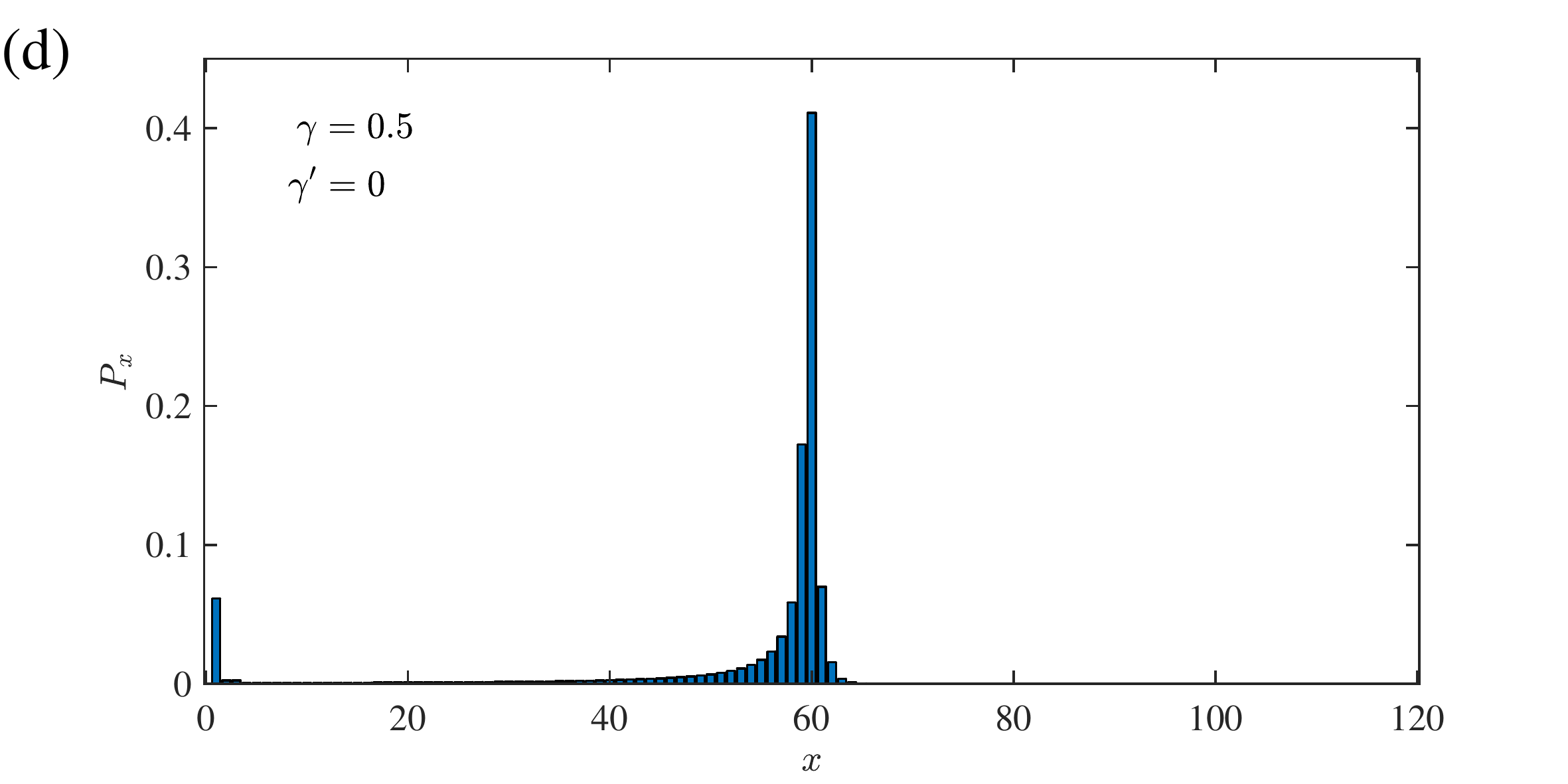}
\caption{(a) The real space Hamiltonian. The local loss probabilities $P_x$ are defined on links (see text). (b) The $D$ matrix. Each box represents a $P_x$ defined on a link [Eq. (\ref{Pdef})]. The diagonal terms $-2(\gamma+\gamma')$, except the first and the last, are shared equally by the two adjacent boxes. (c) Case 1 without edge burst. (d) Case 2 with edge burst. For both (c) and (d), $t=0.8$.}
\label{S3}
\end{figure}

To illustrate the generality of our conclusion, we consider a model whose anti-Hermitian part is $k$-dependent:
\bea
\text{Model III}:\ \ H(k)=2t\text{cos}k+i\gamma\text{sin}k+i\gamma'\text{cos}k-i(\gamma+\gamma'),
\eea
in which the anti-Hermitian part reads
\bea
D(k)=\frac{H(k)-H^\dagger(k)}{2i}=\gamma\text{sin}k+\gamma'\text{cos}k-(\gamma+\gamma').
\eea   The corresponding real space Hamiltonian is shown in Fig.~\ref{S3}(a). The wavefunction norm evolves as
\bea
\frac{d}{dt}\la\psi(t)|\psi(t)\ra &=&i\la\psi(t)|(H^\dagger-H)|\psi(t)\ra\nn\\
&=&2\la\psi(t)| D|\psi(t)\ra\nn\\
&=&\sum_{i,j}2D_{i,j}\la\psi(t)|i\ra\la j|\psi(t)\ra\ra\nn\\
&=&\sum_{i,j}2\psi_i^*(t)D_{i,j}\psi_j(t),
\eea
in which the expression of $D$ matrix is shown in Fig.~\ref{S3}(b). Unlike the model considered in our main article, both on-site terms and hopping terms in $D$ contribute to the particle loss. \\

There is certain freedom in defining the local loss probability $P_x$ in models with $k$-dependent anti-Hermitian part. While the on-site loss can be ascribed to a single site, the loss from a link could be ascribed either to its left or right site. Such freedom is unimportant for our results because different definitions give the same qualitative behavior.  We choose to define $P_x$ at links [Fig.~\ref{S3}(a)] and take
\be
P_x=\int_0^\infty  dt\big[(-\gamma'+i\gamma)\psi^*_x(t)\psi_{x+1}(t)-(\gamma'+i\gamma)\psi_{x+1}^*(t) \psi_x(t)\big] +
\left\{
\begin{aligned}
&(\gamma+\gamma')\int_0^\infty dt \big(2|\psi_x(t)|^2+|\psi_{x+1}(t)|^2\big), \quad x=1 &\\
&(\gamma+\gamma')\int_0^\infty dt \big(|\psi_x(t)|^2+|\psi_{x+1}(t)|^2\big), \quad x\in[2,L-2] &\\
&(\gamma+\gamma')\int_0^\infty dt \big(|\psi_x(t)|^2+2|\psi_{x+1}(t)|^2\big), \quad x=L-1,&
\end{aligned}
\right.\label{Pdef}
\ee
which is pictorially shown in the yellow boxes in Fig.~\ref{S3}(b). Each box in Fig.~\ref{S3}(b) contains a positive semidefinite matrix, and therefore $P_x$ is always a non-negative real number. Note that $\sum_{x=1}^{L-1} P_x = 1$ is satisfied.

Now we consider two cases:
\begin{enumerate}
\item $\gamma=0,\gamma'\ne0$, NHSE is absent.
\item $\gamma\ne0,\gamma'=0$, NHSE is present.
\end{enumerate}
In both cases, the imaginary gap closes, but only the case 2 exhibits the edge burst [Fig.\ref{S3}(c)(d)]. This example further demonstrates that the imaginary gap closing is not a sufficient condition; it must cooperate with NHSE to generate the edge burst.

\bibliography{dirac}